\documentclass{article}
\usepackage{hyperref}

\usepackage{authblk}

\usepackage{latexsym}
\usepackage{graphicx}
\usepackage{amsmath}
\usepackage{amsfonts}
\usepackage{epstopdf}
\usepackage[usenames,dvipsnames,svgnames,table]{xcolor}
\usepackage{xcolor}
\usepackage{color}
\usepackage{setspace} 
\usepackage{natbib}
\usepackage{algorithm,algorithmicx}
\usepackage{algpseudocode}
\usepackage{mathdots}
\usepackage{graphicx} 
\usepackage{enumerate}
\usepackage{relsize}
\usepackage[titletoc]{appendix}
\makeatletter
\newif\if@secnum
\definecolor{lgray}{rgb}{0.2, 0.2, 0.4}
\definecolor{blue1}{rgb}{0.0, 0.4, 0.85}
\definecolor{bluevl}{rgb}{0.94, 0.95, 1}
\definecolor{mDarkTeal}{HTML}{23373b}
\definecolor{white1}{rgb}{0.95, 0.95, 0.96}
\definecolor{mLightGreen}{HTML}{14B03D}
\definecolor{blue2}{rgb}{0.5, 0.0, 0.75}
\definecolor{darkseagreen}{rgb}{0.9, 1, 0.75}
\hypersetup{
	colorlinks =true,
	citecolor = blue1,
	linkcolor = lgray,
	urlcolor = RoyalBlue}

\newtheorem{remark}{Remark}

\newcommand{\xh}{\textbf{x}} 

\numberwithin{equation}{section}
\begin{document}
		\title{Modeling of depth-induced wave breaking in a fully nonlinear free-surface potential flow model}
	\author[1,2]{Ch.E. Papoutsellis \thanks{Corresponding Author: cpapoutsellis@gmail.com} }
	\author[2]{M. Yates}
	\author[2]{B. Simon}
	\author[1]{M. Benoit}
	\affil[1]{Aix Marseille Univ, CNRS, Centrale Marseille, Institut de Recherche sur les Phénomènes Hors-Equilibre (IRPHE), UMR 7342, Marseille, France}
	\affil[2]{Saint-Venant Hydraulics Lab., Cerema, Chatou, France}
\date{}
\maketitle	
\begin{abstract}
	Two methods to treat wave breaking in the framework of the Hamiltonian formulation of free-surface potential flow are presented, tested, and validated. The first is an extension of \cite{Kennedyetal2000}'s eddy-viscosity approach originally developed for Boussinesq-type wave models. In this approach, an extra term, constructed to conserve the horizontal momentum for waves propagating over a flat bottom, is added in the dynamic free-surface condition. In the second method, a pressure distribution is introduced at the free surface that dissipates wave energy by analogy to a hydraulic jump \citep{GG2001}. The modified Hamiltonian systems are implemented using the Hamiltonian Coupled-Mode Theory, in which the velocity potential is represented by a rapidly convergent vertical series expansion. Wave energy dissipation and conservation of horizontal momentum are verified numerically. Comparisons with experimental measurements are presented for the propagation of a breaking dispersive shock wave following a dam break, and then incident regular waves breaking on a mildly sloping beach and over a submerged bar. 
\end{abstract}

\noindent\textbf{Keywords:}
	wave breaking, Hamiltonian formulation of water waves, eddy-viscosity, fully nonlinear water waves\\
	\begin{center}\textbf{Highlights}
		\begin{itemize}
			\item Two wave breaking techniques are tested in a fully nonlinear model. 
			\item Breaking is implemented by extending the Hamiltonian Coupled-Mode Theory. 
			\item Energy dissipation and horizontal momentum conservation are verified numerically. 
			\item Good performance is shown in the presence of nonlinearity, dispersion and breaking. 
			\item Both methods can be implemented in other free-surface potential flow models. 
		\end{itemize}
	\end{center}

	\tableofcontents
	\section{Introduction}
	The continuous rise of human activity in coastal areas significantly motivates the development of numerical models able to predict complex nearshore wave dynamics. Accurate knowledge of evolving wave fields is particularly useful in the study of sediment transport and beach erosion, and in the design of marine renewable energy systems. Challenging issues that must be considered include strongly nonlinear wave propagation, dispersive effects, and complex wave-seabed interactions. In addition, a reliable model should provide accurate predictions of breaking waves and their post-breaking evolution.
	
	Physically precise simulations of unsteady wave propagation, including breaking, can be achieved by solving the primitive Navier-Stokes equations. Although possible, this approach has a high computational cost and its application for large spatial domains (e.g. scales of kilometeres) is cumbersome. Therefore, a widely used approach for simulating coastal waves is the adoption of the physical framework of free-surface potential flow (FSPF) and the derivation of model equations via asymptotic arguments or depth integration. This line of work is mainly represented by Boussinesq-type (BT) models (see e.g. \citet{Nwogu93,CM2009,MFW2006}) and the reviews of \citet{B13,K16}), and Green-Nagdhi (GN) equations (see e.g. \citet{LB09,CLM11,ZDE14,MSM15}). Another class of models that use pertubative calculations is based on the High-Order Spectral (HOS) method \citep{DY87,CS,GN07,GDF2016}. 
	Direct methods of solution of fully nonlinear FSPF equations have also been proposed \citep{BZ07,GAVB14,GGD01}. An alternative approach that treats the fully nonlinear problem is the Hamiltonian Coupled-Mode Theory or HCMT. Like other direct methods of solution, HCMT is a reformulation of FSPF without any simplification concerning nonlinearity, dispersion or seabed deformation. This is achieved through a rapidly convergent vertical series expansion of the velocity potential that is valid in the entire non-uniform fluid domain up to the boundaries \citep{BA11, AP17semi}. The resulting equations retain the dimensionally-reduced Hamiltonian structure of the water wave system \citep{Zak, CS} and give accurate predictions of fully nonlinear and strongly dispersive waves over variable bathymetry up to breaking \citep{Ath15omae,PA17arxiv,ABP17rogue,PCA18}. Wave models with similar mathematical structure and capabilities have also been developed using Chebyshev series representations of the potential in the vertical direction \citep{TS08,YB15,RBY16,RBY19}. The purpose of this paper is to extend the domain of application of fully nonlinear potential flow models by incorporating wave-breaking effects. Here, the modeling strategies are implemented and tested using HCMT.
	
	Most previous work attempts to incorporate wave breaking into BT or GN models by introducing dissipation mechanisms that are applied throughout breaking events. This is consistent with the fact that wave breaking results in wave energy dissipation. This approach thus avoids the direct fine-scale computations of the turbulent wave motion encountered during breaking and parametrizes the effects of wave breaking on the wave kinematics. The simulated decrease in wave energy suppresses overturning of the free surface allowing the wave form (and simulation) to remain stable. Implementation of such an approach requires (i) the addition of an extra dissipative term to the otherwise inviscid evolution equations and (ii) the adoption of a breaking criterion to determine when the dissipative term will be activated (beginning of breaking) and de-activated (end of breaking). Concerning the extra terms, two dominant strategies can be identified: the Eddy Viscosity Model \citep{HH70,Z91,KK92,Kennedyetal2000} and the Surface Roller Model \citep{Sv84,SMD93,MSS97}. 
	
	In the eddy viscosity approach, a diffusion term is added to the momentum equation. It is controlled by an eddy viscosity coefficient that depends on a mixing length parameter that is used to calibrate the approach in comparison to experimental measurements. \citet{CRB10} proposed an improvement by adding an extra ad hoc diffusive term to the mass equation to improve the simulated horizontal asymmetry in the inner surf zone.
	\cite{KvG14} also proposed a variant of \citet{Kennedyetal2000}'s approach applicable to fully dispersive waves described by a high-order Hamiltonian model. Recently, \cite{KR18} proposed an alternative approach in which the eddy viscosity is calculated in terms of an additional partial differential equation for the turbulent kinetic energy.  Finally, an eddy viscosity approach for deep water breaking using the HOS method was presented in \citet{TPC10} and \citet{SD17}. 
	
	In the second approach, the surface roller concept asserts that a breaking wave is divided into a potential core and a turbulent region close to the wave front. By prescribing horizontal velocity profiles in these regions, an extra term is derived in the momentum equation. Dissipation depends on several geometric parameters, such as the roller thickness and the mean front slope of the breaking wave. Advanced versions of this approach that take into account rotational effects have also been proposed \citep{VS00,B2004}.
	 
	A new Hybrid approach has also gained popularity due to its relative simplicity. In this technique, instead of adding an extra term that dissipates wave energy, BT or GN models are locally transformed into the nonlinear shallow water (NSW) equations by suppression of the dispersive terms. This allows breaking waves (detected by an appropriate criterion) to be propagated as shocks \citep{TP09,RCK10,BBCCL11,TBMCL12,KDS14,KR18}. The advantage of this approach is that no additional ad-hoc dissipative terms are required, thus the breaking dissipation introduced in the model has no free parameters requiring calibration. 
     
     In the framework of fully nonlinear FSPF, the development of dissipation mechanisms that model wave breaking has not yet been studied thoroughly. A notable exception is the boundary element implementation of \citet{GG2001} and \cite{GG2019} in which a dissipative pressure distribution is introduced in the dynamic free-surface boundary condition (FSBC). It is constructed so that the energy dissipated is proportional to the energy dissipated by a hydraulic jump with characteristics similar to those of  the breaking wave.
     
      The goal of this study is to address this lack of progress by implementing two wave breaking techniques in the framework of the Hamiltonian formulation of FSPF. The first is an eddy viscosity approach following \citet{Kennedyetal2000}. The main difficulty here and the primary difference of the current approach relative to previous work is that the dynamic FSBC is an evolution equation of the free-surface velocity potential rather than the (depth-integrated) horizontal velocity used in BT or GN models. The second method uses the dissipative pressure distribution proposed in \citet{GG2001}. The two methods are incorporated in a numerical scheme for HCMT \citep{PCA18}. In principle, the presented techniques can be applied to other numerical models provided that the dynamic FSBC is cast in terms of the free-surface potential (e.g. \citet{BZ07,GAVB14,GGD01,RBY16,GDF2016}).
     
     The organisation of the paper is as follows: in Section \ref{Sec:FSPF}, the governing equations of free surface potential flow and its Hamiltonian structure are outlined. In Section \ref{Sec: Breaking}, the breaking initiation criterion and the two wave breaking dissipation techniques are introduced.
     Section \ref{SecNumMet} presents briefly the HCMT and extends the numerical scheme of \citet{PCA18} to the breaking case studied herein. Numerical verifications and validations are presented in Section \ref{SecNumRes}, before a discussion of the results and conclusions are detailed in Section \ref{Sec:Conclusions}.   
     
     \section{Free-surface potential flow modeling }\label{Sec:FSPF}
	\subsection{Governing equations}
	The FSPF equations describe the irrotational motion of an inviscid, incompressible fluid with a free surface under the influence of gravity. In a two-dimensional Cartesian coordinate system $(x,z)$, with the vertical coordinate $z$ pointing upwards, the fluid occupies a region delimited by the time-dependent free surface at $z=\eta(x,t)$ and the fixed bottom at $z=-h(x)$, $h(x)>0$. The free-surface elevation $\eta$ and bottom  bathymetry $h$ are assumed to be smooth, single-valued functions such that $\eta(x,t)+h(x)>0$, thus excluding overturning waves and the representation of the shoreline. Fluid flow is described in terms of the velocity potential $\Phi(x,z,t)$ and free-surface elevation $\eta(x,t)$ by the equations, e.g. \citep{St}
	\begin{subequations}
		\begin{align}
			\partial_t\eta +\partial_x \eta \partial_x   \Phi      -       \partial _{z}   \Phi   &=0, \,\,\,\,\text{on}\,\, z=\eta(x,t),\label{KinematicC}\\
			\partial_t\Phi+\frac{1}{2}(\partial_x\Phi)^2+\frac{1}{2}(\partial_{z}\Phi)^2+g\eta & =-P_{\text{surf}}, \,\,\,\text{on}\,\, z=\eta(x,t),\label{DynamicC}\\
			\partial_x^2\Phi+\partial_{z}^2\Phi & =0, \,\,\,\, \text{in}\,\, -h(x)\leq z\leq\eta(x,t),\label{Laplace}\\
			\partial_x h \partial_x   \Phi      +       \partial _{z}   \Phi & =0, \,\,\,\,\text{on}\,\, z=-h(x),\label{BottomC}
		\end{align}
		\end{subequations}
	where $g$ is the acceleration of gravity, and $P_{\text{surf}}$ denotes the pressure acting on the free surface. It is also assumed that $\eta$, $\Phi$, and their derivatives vanish at infinity. \cite{Zak} observed that the kinematic and dynamic FSBCs, Eqs. \eqref{KinematicC} and \eqref{DynamicC}, can be written equivalently as a system of evolution equations of the free-surface elevation $\eta   (  x  ,  {\kern 1pt} t  )$ and the trace of wave potential on $z=\eta   (  x  ,  {\kern 1pt} t  )$,
	\begin{align}\label{DirCond}
	\psi   (  x  ,  {\kern 1pt} t  )   : =      \Phi   (  x  ,  {\kern 1pt} z=\eta   (  x  ,  {\kern 1pt} t  )  ,  {\kern 1pt} t  ).
	\end{align}
	This system takes the form
	\begin{subequations}\label{ZCS}
		\begin{align}
		\partial_t\eta &= \mathcal{G}[\eta,h]\psi,\\
		\partial_{t}\psi &=-g\eta -\frac{1}{2}(\partial_{x}\psi)^2+\frac{\big(\mathcal{G}[\eta,h]\psi+\partial_{x}\psi\partial_{x}\eta\big)^2}{2\left(1+
			|\partial_{x}\eta|^2\right)}-P_{\text{surf}},
		\end{align}
	\end{subequations}
	where  ${\rm {\mathcal G}}  {\kern 1pt} [  \eta   ,  h  ] \psi $ is the Dirichlet-to-Neumann (DtN) operator introduced by \cite{CS}. It is defined by 
	\begin{equation} \label{DtNdef}
	{\rm {\mathcal G}} [ \eta  ,  h  ]  {\kern 1pt} \psi       =      -\partial_{x} \eta   {\kern 1pt}   \left[{\kern 1pt} \partial_{x}   \Phi   \right]_{{\kern 1pt} z  {\kern 1pt} ={\kern 1pt}   \eta }     +      \left[{\kern 1pt} \partial _{z}   \Phi   \right]_{{\kern 1pt} z  {\kern 1pt} ={\kern 1pt}   \eta },
	\end{equation}
	where $\Phi$ is determined by the boundary value problem consisting of the Laplace equation \eqref{Laplace} together with the impermeability condition  on the seabed, Eq. \eqref{BottomC}, and the Dirichlet condition, Eq. \eqref{DirCond}. In Section \ref{Sec: Breaking}, specific forms of $P_{\text{surf}}$ that model wave-breaking dissipation will be presented. Prior to that, it is instructive to briefly present the conservation of three important quantities when $P_{\text{surf}}\equiv 0$, namely the total energy, mass and horizontal momentum.

	\subsection{Hamiltonian structure and conserved quantities}
	In the absence of surface pressure, the water wave system \eqref{ZCS} is equivalent with the Hamiltonian formulation \citep{Zak,Broer74,Miles}       
	 \begin{subequations}\label{ZCSHam}
	 	\begin{align}
	 	\partial_t\eta&=\delta_{\psi} \mathcal{H}= \mathcal{G}[\eta,h]\psi,\label{eq:KC}\\
	 	\partial_{t}\psi&= -\delta_{\eta} \mathcal{H}=-g\eta -\frac{1}{2}(\partial_{x}\psi)^2+\frac{\big(\mathcal{G}[\eta,h]\psi+\partial_{x}\psi\partial_{x}\eta\big)^2}{2\left(1+
	 		|\partial_{x}\eta|^2\right)},\label{eq:DC}
	 	\end{align}
	 \end{subequations}
	 	where $\delta_{\eta}:=\delta/\delta\eta$ and $\delta_{\psi}:=\delta/\delta\psi$ denote variational derivatives, and the Hamiltonian $\mathcal{H}=\mathcal{H}[\eta,\psi](t)$ is the functional of the total (kinetic and potential) wave energy 
	 \begin{align}\label{Hamiltonian}
	 \begin{aligned}
	 \mathcal{H}[\eta,\psi](t) = \int  E (\eta,\psi)\, dx,\quad  
	 E(\eta,\psi)=\frac{1}{2}\left(\psi\mathcal{G}[\eta,h]\psi +g\eta^2\right).
	 \end{aligned}
	 \end{align}
The systematic derivation of conserved quantities associated with solutions $(\eta,\psi)$ of the above system is presented in \citet{BO82}. Conservation of energy, mass and momentum are verified explicitly below because these calculations will be used in the next section.
 
 Conservation of energy is a well-known property of the Hamiltonian structure \eqref{ZCSHam},\eqref{Hamiltonian}: the time derivative of $\mathcal{H}(\eta,\psi)$ is
 \begin{align}
	\frac{d\mathcal{H}}{dt} = \int\big(\delta_{\psi}\mathcal{H}\partial_t\psi+\delta_{\eta}\mathcal{H}\partial_t\eta\big)dx=\int\big(-\delta_{\psi}\mathcal{H}\delta_{\eta}\mathcal{H}+\delta_{\eta}\mathcal{H}\delta_{\psi}\mathcal{H}\big)dx=0,
	\end{align}
after using the first equalities of Eqs. \eqref{ZCSHam}. For the \emph{mass}, 
\begin{align}\label{Mass}
\mathcal{M}=\int\eta dx,
\end{align}
the time derivative is
	\begin{align*}
	\frac{d\mathcal{M}}{dt} = \int\partial_t\eta dx = \int G[\eta,h]\psi dx = 0,
	\end{align*}
	where the last equality is obtained after invoking the self-adjointness of the operator i.e. $\mathcal{G}[\eta,h]$, $\int G[\eta,h]u\,v dx = \int G[\eta,h]v\,u dx  $ for all $u$, $v$ \citep{Milder90,Lannesb}, in conjunction with the fact that $\mathcal{G}[\eta,h]1=0$\footnote{Noting that the unique solution of the boundary value problem Eq. \eqref{Laplace}, \eqref{BottomC} and \eqref{DirCond} with $\psi=1$ is $\Phi(x,z) = 1$, the definition of the DtN operator, Eq. \eqref{DtNdef} leads to $G[\eta,h]1=0$.}. Finally, the horizontal momentum in the case of a flat bottom, $\partial_{x}h = 0$, is \citep{BO82} 
\begin{align}\label{eq:impulse}
\mathcal{I}(\eta,\psi)=-\int \partial_{x}\eta\,\psi\, dx.
\end{align} 
	The time derivative of functional $\mathcal{I}$ is thus
		\begin{align}\label{eq:dIdt}
		\begin{aligned}
			\frac{d\mathcal{I}}{dt} = -\int\Big( \partial_{xt}^2\eta\psi + \partial_{x}\eta\partial_{t}\psi \Big) d x
			=-\int\Big( -\partial_t\eta\partial_{x}\psi + \partial_{x}\eta\partial_{t}\psi\Big) d x,
		\end{aligned}
		\end{align}
	where integration by parts has been used for the first term in conjunction with the assumption that the temporal derivative of the free-surface elevation vanishes at infinity. Substituting the Hamiltonian equations \eqref{ZCSHam} in Eq. \eqref{eq:dIdt}, $d\mathcal{I}/dt$ is written 
	\begin{align}\label{eq:VarII}
	\frac{d\mathcal{I}}{dt}=\int\Big( \delta_{\psi}\mathcal{H}\partial_{x}\psi + \delta_{\eta}\mathcal{H}\partial_{x}\eta \Big) d x,
	\end{align}
	and the identity
	\begin{align}\label{eq:IdE}
	\int\partial_{x}E dx = \int\Big(\delta_{\psi}\mathcal{H}\partial_{x}\psi + \delta_{\eta}\mathcal{H}\partial_{x}\eta\Big)dx,
	\end{align}
	 proven in \ref{App:dE_dxH}, yields 
	 \begin{align}
	 d\mathcal{I}/dt = 0,
	 \end{align}
	 since $E$, given by Eq. \eqref{Hamiltonian}, vanishes at infinity. Here it is thus proven that the total mass, energy and horizontal momentum (in the case of uniform water depth) are all constant in time.
	 
	 \section{Wave-breaking models}\label{Sec: Breaking}
	To take into account the effects of wave breaking, the modified water-wave system\footnote{It should be noted that the modified Bernoulli equation \eqref{DCP} has been considered previously when treating wave absorption at the lateral end of numerical wave tanks (see e.g. \citet{CBS93,GH1997}).} is:
	\begin{subequations}\label{ZCSP}
		\begin{align}
		\partial_t\eta&=\delta_{\psi}\mathcal{H},\label{KC}\\
		\partial_{t}\psi&=-\delta_{\eta}\mathcal{H} - P_{\text{surf}},\label{DCP}
		\end{align}
	\end{subequations}
	where $P_{\text{surf}}=P_{\text{surf}}(x,t)$ is activated from the initial time of breaking, $t_i$, until the cessation time of breaking, $t_f$, over a spatial \emph{breaking region} $\Theta$, which also varies in time. The domains $\Theta$ and $[t_i,t_f]$ are determined during the wave evolution on the basis of empirical criteria described in Sections \ref{sub_EVM} and \ref{sub_EHJ}. It should be noted that the expressions of the total energy, mass, and horizontal momentum do not change and are still defined by Eqs. \eqref{Hamiltonian}, \eqref{Mass} and \eqref{eq:impulse}, respectively.   
	
	The effect of $P_{\text{surf}}$ on the rate of change of total energy becomes apparent by calculating the time derivative of the Hamiltonian functional \eqref{Hamiltonian} by taking into account Eqs. \eqref{ZCSP} such that 
    \begin{align*}
    \frac{d\mathcal{H}}{dt} =\int\big(\delta_{\eta}\mathcal{H}\partial_t\eta+\delta_{\psi}\mathcal{H}\partial_t\psi\big)d\xh=\int\big(\left(-\partial_t\psi-P_{\text{surf}}\right)\partial_t\eta+\partial_t\eta\partial_t\psi\big)d\xh,
     \end{align*}
    and thus,
    \begin{align}\label{eq:calc_dHdt}
    \frac{d\mathcal{H}}{dt} = -\int_{\Theta} P_{\text{surf}} \,\partial_t\eta \,dx=-\int_{\Theta} P_{\text{surf}} \,G[\eta,h]\psi\, dx. 
    \end{align}
   The expression for the rate of change of mass remains unchanged and still equals zero. For the rate of change of horizontal momentum of the modified system \eqref{ZCSP}, Eqs. \eqref{eq:dIdt} and \eqref{ZCSP} are used to calculate
    \begin{align}\label{eq:33}
    \frac{d\mathcal{I}}{dt}=\int\Big( \partial_t\,\eta\partial_{x}\psi - \partial_{x}\eta\,\partial_{t}\psi\Big) d x=\int \big(\delta_{\psi}\mathcal{H}\partial_{x}\psi + \delta_{\eta}\mathcal{H}\partial_{x}\eta+P_{\text{surf}}\,\partial_{x}\eta\big)dx.
    \end{align}
    Using Eq. \eqref{eq:IdE}, the first two terms in the right hand side of Eq. \eqref{eq:33} vanish leading to 
    \begin{align}\label{eq:dIdt_1}
    \begin{aligned}
    \frac{d\mathcal{I}}{dt}=\int_{\Theta} P_{\text{surf}}\,\partial_{x}\eta\, dx.
    \end{aligned}
    \end{align}
    Eqs. \eqref{eq:calc_dHdt} and \eqref{eq:dIdt_1} are the starting point for the construction of $P_{\text{surf}}$.

     \subsection{Eddy Viscosity Model \normalfont(EVM)}\label{sub_EVM}
     In the eddy viscosity approach, $P_{\text{surf}}$ is constructed with the requirement that the momentum $\mathcal{I}=\mathcal{I}(\eta,\psi)$, Eq.  \eqref{eq:impulse}, remains invariant over a flat bottom. 
     As noted by \cite{Kennedyetal2000}, this feature of horizontal momentum conservation is consistent with the analogy between breaking bores and breaking waves. Using Eq. \eqref{eq:dIdt_1}, it is not straightforward to determine $P_{\text{surf}}$ to enforce the condition $d\mathcal{I}/dt=0$. Nonetheless, by assuming that $P_{\text{surf}}$ vanishes at the boundary of the breaking region $\Theta$ and that the bottom is flat ($\partial_xh=0$), an integration by parts gives
     \begin{align}\label{eq:varI2}
     \begin{aligned}
     \frac{d\mathcal{I}}{dt}
     =-\int_{\Theta} \partial_x P_{\text{surf}}\,H\, dx,
     \end{aligned}
     \end{align} 
     where $H=\eta+h$ is the total water depth. Following \cite{KvG14}, $P_{\text{surf}}$ is chosen here such that
     \begin{align}\label{eq:Px}
     \partial_x P_{\text{surf}}= \frac{1}{H}\partial_{x}F,\quad\text{with}\quad F=0\quad \text{outside}\,\,\Theta,
     \end{align}
     which yields $d\mathcal{I}/dt=\int_{\Theta}\partial_x F = 0$. Function $F$ is given by
      \begin{align}\label{PVG}
      F = \nu\, G[\eta,h]\psi,
      \end{align}
      where $\nu = \nu(x,t)$ is the eddy viscosity coefficient assumed to be of the form: 
      \begin{align}\label{eq:evc}
      \nu = \delta^2 B(\partial_t\eta) H G[\eta,h]\psi.
      \end{align}
      In the above equation, $\delta$ is a mixing length coefficient that may be used as a calibration parameter. $B(\partial_t\eta)$ is a coefficient that varies smoothly in time from 0 to 1, ensuring numerically the smooth initiation of wave breaking:
      \begin{align}
      B(\partial_t\eta) &= \left\{
      \begin{aligned}
      0,&\quad \partial_t\eta\leq \partial_t\eta^*\\
      \frac{\partial_t\eta}{\partial_t\eta^*}-1,&\quad \partial_t\eta^* <\partial_t\eta< 2 \partial_t\eta^*\\
      1,&\quad \partial_t\eta\geq 2 \partial_t\eta^*
      \end{aligned}\right.
      \end{align}
      where $\partial_t\eta^*$ is a function of time that equals an initial threshold value $\partial_t\eta^{\text{I}}$ and decreases linearly to the final value $\partial_t\eta^{\text{F}}$ from the initial time of breaking $t_{i}$ during a transition time $T^*$: 
      \begin{align}
      \partial_t\eta^* &= \left\{\begin{aligned}
      \partial_t\eta^{\text{I}} - \frac{(t-t_{\text{br}})}{T^*}\left(\partial_t\eta^{\text{I}}-\partial_t\eta^{\text{F}}\right),&\quad t_{i}\leq t\leq t_{i}+ T^*\\
      \partial_t\eta^{\text{F}},&\quad  t\geq t_{i}+ T^*
      \end{aligned}\right.
      \end{align}
      The values for $\partial_t\eta^{\text{I}}$ and $\partial_t\eta^{\text{F}}$ are assumed proportional to the shallow water speed $\sqrt{g h}$: a wave is considered to start breaking when $\partial_t\eta>\partial_t\eta^{\text{I}}:=\gamma_\text{I}\sqrt{g h}$, and the breaking process terminates when $\partial_t\eta>\partial_t\eta^{\text{F}}:=\gamma_\text{F}\sqrt{g h}$. The values of $\gamma_{\text{I}}$ and $\gamma_{\text{F}}$ depend on the examined configuration (bathymetry and type of wave breaking) and are used as calibration parameters to minimize the difference between the numerical results and the experimental measurements. The transition time $T^*$ is fixed as $T^*=5\sqrt{h/g}$, as suggested by \citet{Kennedyetal2000}. The breaking region $\Theta$ spans the front of the wave, such that $\Theta=[x_c(t),x_r(t)]$ where $x_c(t)$, $x_r(t)$ are the abscissae of the crest and the following trough, respectively.
      
\begin{remark}
	The main difficulty in the implementation of the EVM is that Eq. \eqref{eq:Px} determines the spatial derivative $\partial_xP_{\text{surf}}$, while $P_{\text{surf}}$ itself is needed in the dynamic FSBC \eqref{DCP}. Note that this is not an issue in BT models since they are derived using the horizontal derivative of Eq. \eqref{DCP}. It is shown here (Section \ref{subsubEVM}) how $P_{\text{surf}}$ can be computed.
\end{remark}       

\subsection{Effective Hydraulic Jump model \normalfont(EHJ)}\label{sub_EHJ}

A simple analysis of Eq. \eqref{eq:calc_dHdt} suggests that any expression of the form $P_{\text{surf}}=\lambda\mathcal{G}[\eta,h]\psi$ with $\lambda>0$ results in wave energy dissipation:  $d\mathcal{H}/dt=-\lambda\int_{\Theta}  \,(G[\eta,h]\psi)^2\, dx<0$. Elaborating on this idea, \cite{GG2001} proposed the expression
\begin{align}\label{EHJ}
P_{\text{surf}}(x,t)  = \nu(t)\,S(x)\, \mathcal{G}[\eta,h]\psi,
\end{align}
where $S(x)$ is a (smooth) bump function supported in the breaking region and $\nu = \nu(t)$ is a time dependent function that determines the characteristics of breaking and must be calculated during the wave evolution.

\begin{figure}
	\centering
	\includegraphics[scale = 0.5]{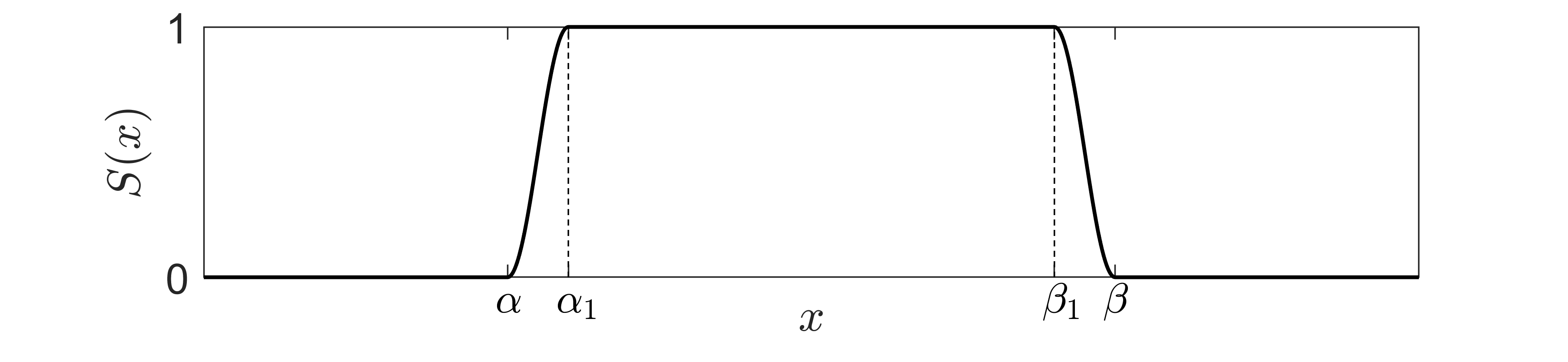}
	\caption{The bump function $S(x)$}
	\label{fig:funS}
\end{figure}

The function $S$ ensures a smooth transition in space between breaking and non-breaking regions. Assuming that the breaking region is the interval $\Theta=[\alpha,\beta]\subseteq [x_l,x_r]$,
and introducing the internal points $\alpha_1 = \alpha+\sigma|\alpha-\beta|$, $\beta_1 = \beta-\sigma|\alpha-\beta|$, the function $S$ used in this work is written
\begin{align}
S = \left\{\begin{aligned}
0,&\quad x\leq \alpha\\
-\frac{1}{2}\cos\left(\pi\frac{x-\alpha}{\alpha_1-\alpha}\right)-\frac{1}{2},&\quad \alpha\leq x\leq \alpha_1\\
1,&\quad \alpha_1\leq x\leq \beta_1\\
\frac{1}{2}\cos\left(\pi\frac{x-\beta_1}{\beta_1-\beta}\right)+\frac{1}{2},&\quad \beta_1\leq x\leq \beta\\
0,&\quad x\geq \beta
\end{aligned}\right.
\end{align}
and is depicted in Figure \ref{fig:funS}. The parameter $\sigma$ controls the transition length of $S(x)$ between 0 and 1 at both ends of the breaking region. The value $\sigma= 0.1$ is used in all of the test cases presented here. Smaller values of $\sigma$ should be avoided because they increase the slope of $S$ near the boundaries of $\Theta$, $|\partial_xS|\leq \pi/(2|\alpha-\beta|\sigma)$ and may cause instabilities.  

\begin{figure}
	\begin{center}
		\includegraphics[scale = 0.6]{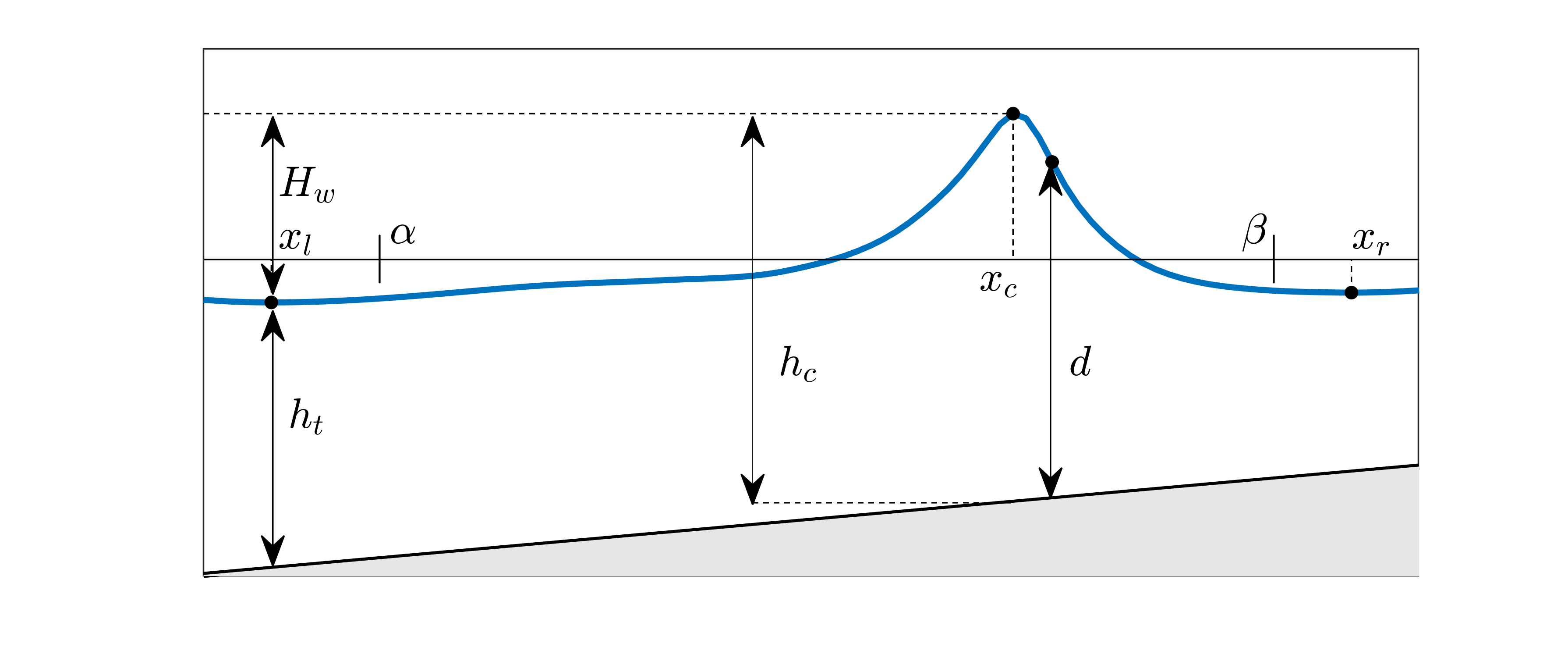}
		\caption{Geometric parameters used in the EHJ model.}
		\label{fig:geometry}
	\end{center}
\end{figure}

In order to estimate $\nu(t)$, \citet{GG2001} asserted the similarity between a breaking wave and a hydraulic jump, proposing the following strategy: assume that the instantaneous power dissipated by $P_{\text{surf}}$, defined as $\Pi_{\text{surf}}$, is of the same order as the instantaneous power dissipated by a hydraulic jump, $\Pi_{\text{h}}$. Then, $\Pi_{\text{surf}}$ may be expressed as 
\begin{align}\label{eq:effhb}
\Pi_{\text{surf}}=\nu_0 \Pi_{\text{h}},
\end{align}
where $\nu_0$ is a positive calibration constant. 
Eq. \eqref{eq:effhb} will then be used to calculate $\nu(t)$ at every time $t$. Therefore, $\Pi_{\text{surf}}$ and $\Pi_{\text{h}}$ have to be expressed in terms of known quantities and $\nu(t)$. $\Pi_{\text{surf}}$ is given by the integral
\begin{align}
\Pi_{\text{surf}} &= \int P_{\text{surf}}\,G[\eta,h]\psi\,\sqrt{(\partial_x\eta)^2+1}dx,
\end{align}
which after substitution of Eq. \eqref{EHJ} becomes
\begin{align}\label{PsurfEHJG}
\Pi_{\text{surf}}&=\nu(t)\int_{\Theta}S\,(G[\eta,h]\psi)^2\,\sqrt{(\partial_x\eta)^2+1}dx.
\end{align}
For the calculation of $\Pi_{\text{h}}$, \cite{GG2001} proposed the following formula 
\begin{align}\label{PEHJ}
\Pi_{\text{h}} =  g\, c\, d\, \frac{H_{\text{w}}^3}{4 h_c h_l},
\end{align}      
where $c$ is the phase speed, $d$ is the depth below the point of maximum front slope, $H_{\text{w}}(t)=\eta(x_c,t)-\eta(x_l,t)$ is the wave height defined as the height between the crest and the previous trough, $h_c(t) = \eta(x_c,t)+h(x_c)$ is the depth below the crest and $h_t = \eta(x_l,t)+h(x_l)$  is the depth below the trough (see Figure \ref{fig:geometry}).
The breaking initiation and termination criterion are the same as in the EVM approach. Concerning the breaking region $\Theta$, two different possibilities will be examined: (i) the region proposed by \citet{GG2001} in which the breaking region extends from the crest to two points $x=\alpha$ and $x=\beta$ on each side of the crest (see Figure \ref{fig:geometry}):
\begin{align}\label{eq:bzGG}
\Theta:=[\alpha,\beta]=\left\{x\in[x_l,x_r]:\left|\frac{\partial_n\Phi}{\max\partial_n\Phi}\right|> \varepsilon\quad \text{and}\quad \left|\frac{\partial_n\Phi}{\min\partial_n\Phi}\right|> \varepsilon\right\}, 
\end{align}
where $\partial_n\Phi$ is the normal velocity on the free surface ($\partial_n\Phi=G[\eta,h]\psi$), $\max \partial_n\Phi$ (resp. $\min \partial_n\Phi$) is the maximum (resp. minimum) of $\partial_n\Phi$ in the interval $[x_l,x_r]$, and $\varepsilon$ is a small threshold value ($\varepsilon = 10^{-4}$ in this work), and (ii) a breaking region covering only the front face of the wave, spanning from the breaking crest to the following trough, $\Theta=[x_c,x_r]$, as in the EVM. The above two options will be refereed to as EHJ and EHJ$_f$, respectively. Note that the requirement that $d\mathcal{H}/dt$ is negative is not affected by the extent of the breaking region $\Theta$.

	\section{Numerical implementation}\label{SecNumMet}
	\subsection{Hamiltonian Coupled-Mode Theory}
	The numerical implementation of Eqs. \eqref{ZCSP} is performed by using the Hamiltonian Coupled-Mode Theory (HCMT) \citep{Ath15omae,PA17arxiv}, which is briefly described below. In this approach, the unknown velocity potential $\Phi(x,z,t)$ is expressed as a rapidly convergent series expansion in terms of unknown horizontal functions $\{\varphi_n\}_{n=-2}^{\infty}=\{\varphi_n(x,t)\}_{n=-2}^{\infty}$ and prescribed vertical basis functions $\{Z_n\}_{n=-2}^{\infty}=\{Z_n(z;\eta(x,t),h(x))\}_{n=-2}^{\infty}$:
	\begin{align}\label{expansion}
	\Phi(x,z,t) = \sum\limits_{n=-2}^{\infty}\varphi_n(x,t)Z_n(z;\eta(x,t),h(x)).
	\end{align} 
	The exact form of $\{Z_n\}_{n=-2}^{\infty}$ is given in \ref{App:Zn} by Eqs. \eqref{eq:Zadda}, \eqref{eq:Zaddb}, \eqref{Zntrigc} and \eqref{Zntrigd}. More information about the above expansion and its convergence can be found in \citet{AP17semi}. Using expansion \eqref{expansion} and restricting the horizontal computational domain to the interval $[a,b]$, Eqs. \eqref{ZCSP}  take the form \citep{PA17arxiv}
	\begin{subequations}\label{HCMS}
		\begin{align}
		\partial _{{\kern 1pt} t} \eta    & =    -    (  \partial _{x}^{} \eta   ) (  \partial _{x}^{} \psi   )    +    (  {\kern 1pt} |\partial _{x}^{} \eta {\kern 1pt}   |^{  2}   +  1  )    \left(  h_{  0}^{  -    1}   {\kern 1pt} \mathcal{F}[\eta,h]\psi     +    \mu _{  {\kern 1pt} 0}   \psi   \right),\\
		\partial _{{\kern 1pt} t} \psi    & =    -    g{\kern 1pt} \eta     -    \frac{1}{2} (  \partial _{x}^{} \psi   )^{  2} +    \frac{1}{2} ({\kern 1pt}   |\partial _{x}^{} \eta {\kern 1pt}   |^{  2}   +  1  )    \left(  h_{  0}^{  -   1}  {\kern 1pt} \mathcal{F}[\eta,h]\psi  + \mu _{ 0}  \psi  \right)^{ 2} +P_{\text{surf}},
		\end{align}
	\end{subequations}
	where $\mathcal{F}[\eta,h]\psi     =    \varphi _{-2} (  x  ,  {\kern 1pt} t  )$ is the first element of the sequence that solves the following coupled-mode system
	\begin{subequations}\label{CMS}
		\begin{align}
		\sum _{n    =    -2}^{\infty }\left(A_{mn}^{} \partial _{x}^{2}     +    B_{mn}^{}  \partial _{x}^{}     +    C_{mn}^{} \right)  \varphi _{n}     &=      0  ,\quad                m\ge -2,\quad a<x<b,\label{CMSa}\\
		\sum _{n    =    -2}^{\infty }{\kern 1pt} \varphi _{n}       & =      \psi,\quad a<x<b,\label{CMSb}\\
		\sum _{n    =    -2}^{\infty }\left(A_{m  n}^{} \partial _{x}     +   \frac{1}{2} B_{m  n} \right)  \varphi _{n}     &=      g_m  ,\quad                m\ge -2, \quad x=a,\label{BC1}\\
		\sum _{n    =    -2}^{\infty }\left(A_{m  n}^{} \partial _{x}     +   \frac{1}{2} B_{m  n} \right)  \varphi _{n}     &=      0  ,\quad                m\ge -2, \quad x=b,\label{BC2}
		\end{align}
	\end{subequations}
with
\begin{subequations}
	\begin{align}
	A_{mn}  & = \int_{-h}^{\eta} Z_n Z_m dz=A_{nm},\\
	B_{mn}& = 2\int_{-h}^{\eta}\partial_{x}Z_n Z_m dz + \partial_{x}h\big[Z_n Z_m\big]_{z=-h},\\
	C_{mn}  & =\int_{-h}^{\eta}\Delta Z_n Z_m dz +\partial_{x}h\big[\partial_{x} Z_n Z_m\big]_{z=-h}-\big[\partial_{z} Z_n Z_m\big]_{z=-h}.
	\end{align}
\end{subequations}
	The $(  x  ,  t  )-$dependent matrix coefficients $A_{m  n}^{} =A_{m n}( \eta, h )$, $B_{m  n}^{} =B_{m  n}(  \eta   ,  h  )$ and $C_{m  n}^{} =C_{m  n}^{} (  \eta   ,  h  )$ are  calculated analytically in terms of $\eta   (  x  ,  t  )$, $h  (  x  )$ and the numerical parameters $h_0$, $\mu_0$ that are involved in the definition of $\{Z_n\}_{n=-2}^{\infty}$ \cite[Section 4]{PCA18}. The boundary condition expressed by Eq. \eqref{BC1} forces the horizontal fluid velocity at the vertical section $x=a$ to match
	\begin{align}
	g_m(t)=\int_{h(a)}^{\eta(a,t)}\partial_x\Phi(a,z,t)Z_m(z)dz,
	\end{align}
	and it is used to generate incident waves (Sections \ref{SubsecTK} and \ref{SubsecBB}). If $g_m=0$, the boundary condition Eq. \eqref{BC1} represents a reflecting wall, such as in Eq. \eqref{BC2}.  
	  As opposed to the standard definition of the DtN operator $G[\eta,h]\psi $ of the Laplace problem in the entire fluid domain, $\mathcal{F}[\eta,h]\psi $ is determined at every time $t$,
	by solving a system of differential
	equations in the fixed horizontal domain $[a,b]$. In fact, as shown in \citet{AP17semi}, $G[\eta,h]\psi $ and $\mathcal{F}[\eta,h]\psi $ are related by the following formula
	\begin{equation} \label{DtNrep}
	{\rm {\mathcal G}}   [  \eta   ,   h  ]   \psi       =      -    (  \partial_{x} \eta   )     (  \partial_{x} \psi   )    +    \left(   (\partial_{x}\eta) ^{  2}   +  1  \right)    \left(  h_{  0}^{  -    1}    \mathcal{F}[\eta,h]\psi       +     \mu _{ 0}   \psi   \right).
	\end{equation}
	 
		\subsection{Computation of the DtN Operator}
		In order to compute the DtN operator, the coupled-mode system, Eqs. \eqref{CMSa}-\eqref{BC2}, is truncated at a finite order $M$. Horizontal spatial gradients are approximated using a fourth-order finite-difference method applied on a grid $x_i$, $i=1,...,N_X$, of uniform spacing $\delta x$. First and second-order horizontal derivatives of $\varphi_n$, $n=-2,...,M$ are approximated by using the formulae \eqref{fd1} and \eqref{fd2} given in \ref{App:FD}, and a linear system of algebraic equations is formed of the local values $\varphi_n^i$, $n=1,...,N_{\text{tot}}=M+3$, $i=1,...,N_{X}$ (see \citet[Appendix D]{PCA18}). The local values $\mathcal{F}[\eta,h]\psi $ are then recovered and  used in Eq. \eqref{DtNrep} for the computation of the DtN operator. A detailed study of the convergence and accuracy of the above scheme with respect to $N_{\text{tot}}$ can be found in \citet{AP17semi}.

	\subsection{Computation of $P_{\text{surf}}$}
	\subsubsection{Eddy Viscocity Model}\label{subsubEVM}
	With $F$ given by Eq. \eqref{PVG}, Eq. \eqref{eq:Px} is written
	\begin{align}\label{eq:Pode1}
	\begin{aligned}
	\partial_x P_{\text{surf}}&=\delta^2B(\partial_t\eta)\frac{1}{H}\partial_x\left(H\, (G[\eta,h]\psi)^2\right)\\
	&=\delta^2B(\partial_t\eta)\left[\frac{\partial_x H}{H}(G[\eta,h]\psi)^2 + 2\,G[\eta,h]\psi\, \partial_x (G[\eta,h]\psi)\right].
	\end{aligned}
	\end{align}
	 Differentiating Eq. \eqref{eq:Pode1} yields a second-order differential equation
	\begin{multline}\label{eq:Pode3}
	 \partial_x^2 P_{\text{surf}}= \delta^2B(\partial_t\eta)\left[\left(\frac{\partial_x^2 H}{H}-\frac{(\partial_x H)^2}{H^2}\right)\,(G[\eta,h]\psi)^2 + 2\,(\partial_x(G[\eta,h]\psi))^2\right.\\ \left.+2\,G[\eta,h]\psi\left(\frac{\partial_xH}{H}\partial_x(G[\eta,h]\psi)+\partial_x^2(G[\eta,h]\psi)\right)\right].
	 \end{multline}
	 When Eq. \eqref{eq:Pode3} is combined with the boundary conditions
	 \begin{align}\label{eq:Podebcs}
	 P_{\text{surf}}(x_c) = P_{\text{surf}}(x_r) = 0,
	 \end{align}
	 which ensure that $P_{\text{surf}}$ vanishes at the endpoints of the breaking region, a solvable boundary value problem is defined for $P_{\text{surf}}$. At every time $t$, Eqs. \eqref{eq:Pode3} and \eqref{eq:Podebcs} are solved using a fourth-order finite-difference method in accordance with the numerical scheme used for the computation of the DtN operator. The derivatives appearing in the right hand side of Eq. \eqref{eq:Pode3} are calculated using formulae \eqref{fd1} and \eqref{fd2}.  
	 
\subsubsection{Effective Hydraulic Jump}
The geometrical characteristics appearing in Eq. \eqref{PEHJ} are easily calculated during the wave evolution. However, the estimation of the phase speed $c$ is not straightforward since breaking waves do not have a permanent form. Therefore, to calculate the phase speed, the spatial Hilbert transform technique proposed by \citet{KvG14} is used. The integral in Eq. \eqref{PsurfEHJG} is calculated using the trapezoidal rule.  
\subsection{Time stepping scheme}
Having established methods to compute $\mathcal{F}[\eta,h]\psi$ (or $G[\eta,h]\psi$) and  $P_{\text{surf}}$, the local values of $(G[\eta,h]\psi)^{i }$ and $P_{\text{surf}}^i$, $i=1,...,N_{X} $, can be used to step forward in time  Eqs. \eqref{HCMS}. The derivatives appearing on the right hand sides of Eqs. \eqref{HCMS} are approximated by the finite-difference formulae \eqref{fd1}. In the case of an initial value problem, the initial conditions $\eta(x,0)$ and $\psi(x,0)$ are propagated using the classical fourth-order four-step Runge-Kutta method. For wave problems involving incident wave conditions, Eqs. \eqref{HCMS} are supplemented by the following expression: 
\begin{align}\label{HCMSnum}
\partial _{t} U  = N( {\kern 1pt} U  )  + L(U)
\end{align} 
where $U=(\eta,\psi)^{  {\rm T}}$, $N(U)$ is the vector function defined by the right-hand sides of Eqs. \eqref{HCMS} and $L(U)=c_a(U-U_{\text{in}})+c_b(0\,, G[\eta,h]\psi)^{{\rm T}}$. Functions $c_a(x)$ and $c_b(x)$ are polynomials supported in small regions (sponge layers) before the boundaries $x = a$ and $x = b$ (one and two wavelengths of the incident wave for $c_a$ and $c_b$, respectively). The first term of $L(U)$ is used for wave generation and absorption: $U_{\text{in}}$ denotes the desired incident wave condition, such that $U_{\text{in}}=D(t)(\eta_{\text{in}},\psi_{\text{in}})^{  {\rm T}}$, where $(\eta_{\text{in}},\psi_{\text{in}})^{  {\rm T}}$ is a steady travelling wave solution of system \eqref{HCMS} \cite[Chapter 6]{Papoutsellis}, and $D(t)$ is a smooth function ranging from 0 to 1 in a time interval of typically 3-5 incident wave periods. The second term implies a pressure-type wave absorption.  

A routine is applied at each time step to evaluate the breaking criterion at every node. The wave characteristics (crest and trough locations) are saved at every time step and are passed to the following time step in order to enable tracking of breaking and non-breaking waves. The above scheme requires four solutions of the linear discretized system \eqref{CMS} and one application of the wave tracking algorithm per time step. It is implemented by using \textsc{Matlab\textsuperscript\textregistered}   on an IntelCore i5-8300H, 2.30 GHz. The CPU time needed for the wave tracking algorithm is a small percentage of the total CPU time per time step. Depending on the discretization used (number of modes $N_{tot}$  and number of grid points $N_X$) this percentage ranges from 5-10 \% of the total CPU time per time step. The difference in CPU time between the two proposed methods is insignificant. For the three test cases studied in this paper, the total CPU times per time step are approximately 0.11 sec (Section \ref{SubsecDSW}), 0.1 sec (Section \ref{SubsecTK} ), and 0.23 sec (Section \ref{SubsecBB}).

	\section{Application and validation cases}\label{SecNumRes}
	In this section, the proposed numerical scheme is implemented for the simulation of breaking waves. The optimal set of numerical and breaking parameters is chosen after performing a series of numerical simulations to obtain a good compromise between numerical convergence and the agreement with experimental measurements. To choose the optimal number of total modes $N_{\text{tot}}$, a series of simulations are performed by increasing $N_{\text{tot}}$ (starting with $N_{\text{tot}}=4$) until numerical convergence of the free-surface elevation is achieved. Before presenting the simulation results, the calibration process of both breaking models is described. For a given test case, the parameters controlling the initiation ($\gamma_{\text{I}}$) and termination ($\gamma_{\text{F}}$) of wave breaking are the same for each wave breaking approach. The value of these two parameters depends on the configuration (e.g. bathymetry)  and type of wave breaking studied. The EVM approach has two free parameters, namely $\delta$ and $T^*$, which were set to $\delta=1.5$ and $T^*=5\sqrt{h/g}$ in all cases. The EHJ approach has three free parameters: the coefficient $\nu_0$, the threshold value $\varepsilon$ that determines the length of the breaking region, and the factor $\sigma$ appearing in the definition of the smooth function $S(x)$. The simulation results were not very sensitive to the value of the two latter parameters, which were thus set to $\varepsilon=10^{-4}$ and $\sigma=0.1$ in all simulations. For the coefficient $\nu_0$, the value $\nu_0=1.6$ was chosen in the case of a sloping bottom and $\nu_0=1.0$ in the case of a flat bottom. These parameters remained identical for the EHJ and EHJ$_f$ simulations. 
		
%

	\subsection{Dispersive shock waves}\label{SubsecDSW}
	In this section, the propagation of two breaking dispersive shock waves (DSW) are simulated over a flat bottom. The first test aims to verify the dissipation and conservation properties of the proposed methods, and the second  test reproduces a laboratory experiment.
	\subsubsection{Momentum conservation and energy dissipation}
	\begin{figure}
		\centering
		\includegraphics[scale=0.7]{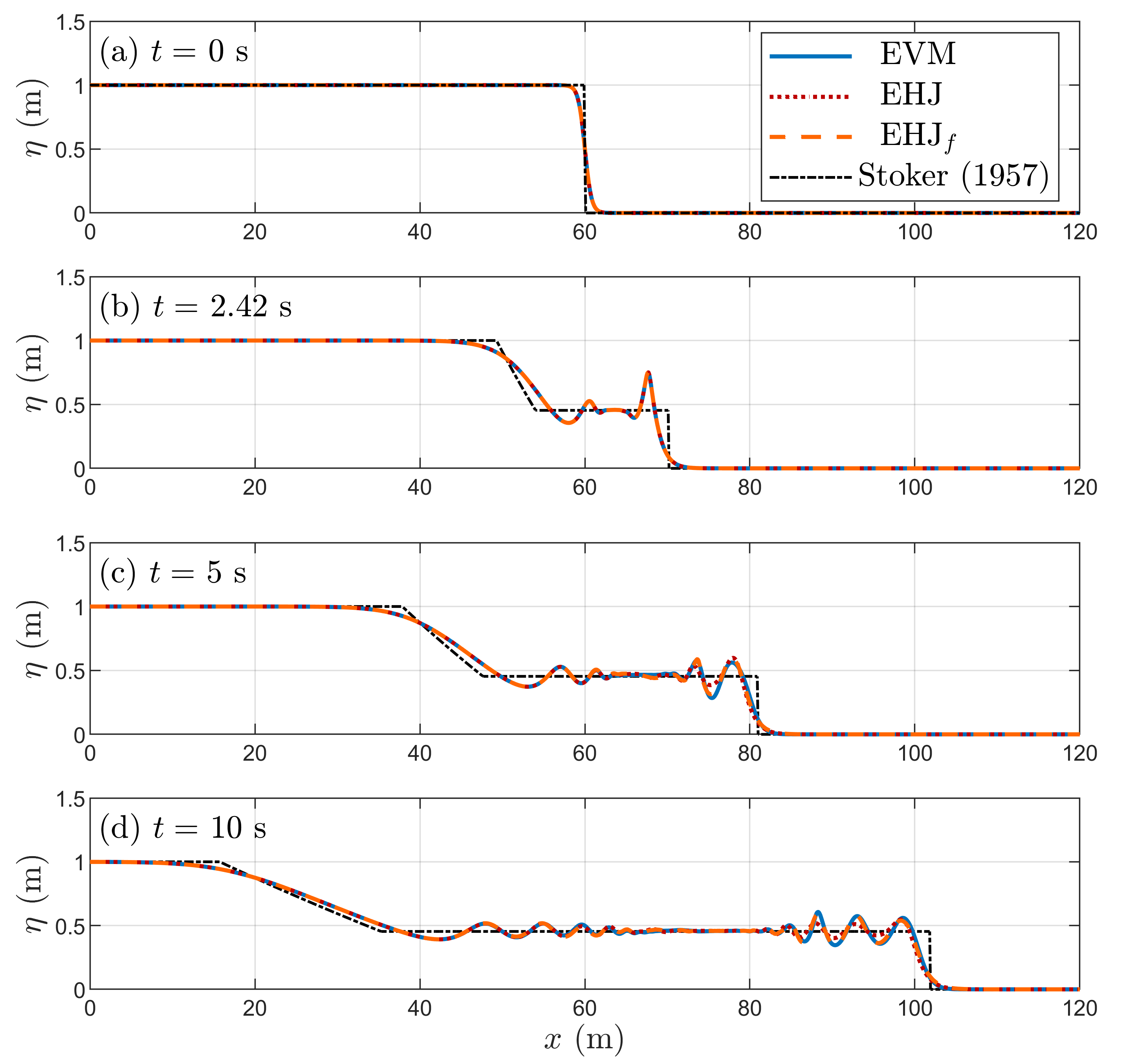}
		\caption{Snapshots of the free-surface elevation for the case of a breaking DSW.}
		\label{Fig_snap}
	\end{figure}
	Two important questions are whether the EHJ method, constructed by requiring energy dissipation, conserves momentum, and whether the EVM, constructed by requiring momentum conservation,  dissipates energy. In order to answer these questions, a test case is presented showing the evolution of two counter-propagating DSWs created after a dam break \citep[see e.g.][]{MITSOT2017}. The horizontal domain extends from -120 to 120 m, and the initial hump of water, centered at $x=0$, is given by  
		\begin{align}
	\eta(x,0)=\frac{1}{2}(H_2-H_1)\left(1+\tanh\left(\frac{\lambda-|x|}{s}\right)\right),
	\end{align}
	with $H_1=1$ m, $H_2=2$ m and $\lambda=60$ m. The parameter $s$ controls the initial surface slope and is chosen as $s=0.8 $ m. The depth measured from $z=0$ is $h=1$ m. The initial free-surface potential is $\psi(x,0)=0$, which implies that the initial fluid velocity and momentum are zero. The simulation is performed using $N_{\text{tot}}=5$ vertical functions and a spatio-temporal discretization $\delta x=0.1$ m, $\delta t=0.01$ s. Wave breaking is initiated using $\gamma_I=0.6$. The value $\gamma_{\text{F}}=0.2$ is used for the termination criterion, but it is not reached during the duration of the simulation. For the EVM, $\delta=1.5$ and for the EHJ, $\nu_0=1$. Snapshots of the simulated free-surface elevations are shown in Figure \ref{Fig_snap} for the right half of the (symmetric) computational domain. For reference, the analytical solution of the NSW equations with a piecewise constant initial condition (Riemann problem) is also plotted (\citet{St}, Section 4.1.1); see also (\citet{NSWstoker_sol}, Section 4.1.1). In the simulations, breaking occurs at $t=2.42$ s, and only the first counter-propagating waves break. The EVM and EHJ$_f$ show nearly identical results while EHJ predicts slightly smaller wave heights for the dispersive tails following the breaking fronts. When comparing to the NSW solution, it is observed that the neglect of dispersive effects in the NSW equations leads to an inadequate description of the DSW's surface elevation. 
	
\begin{figure}
		\centering
		\includegraphics[scale=0.65]{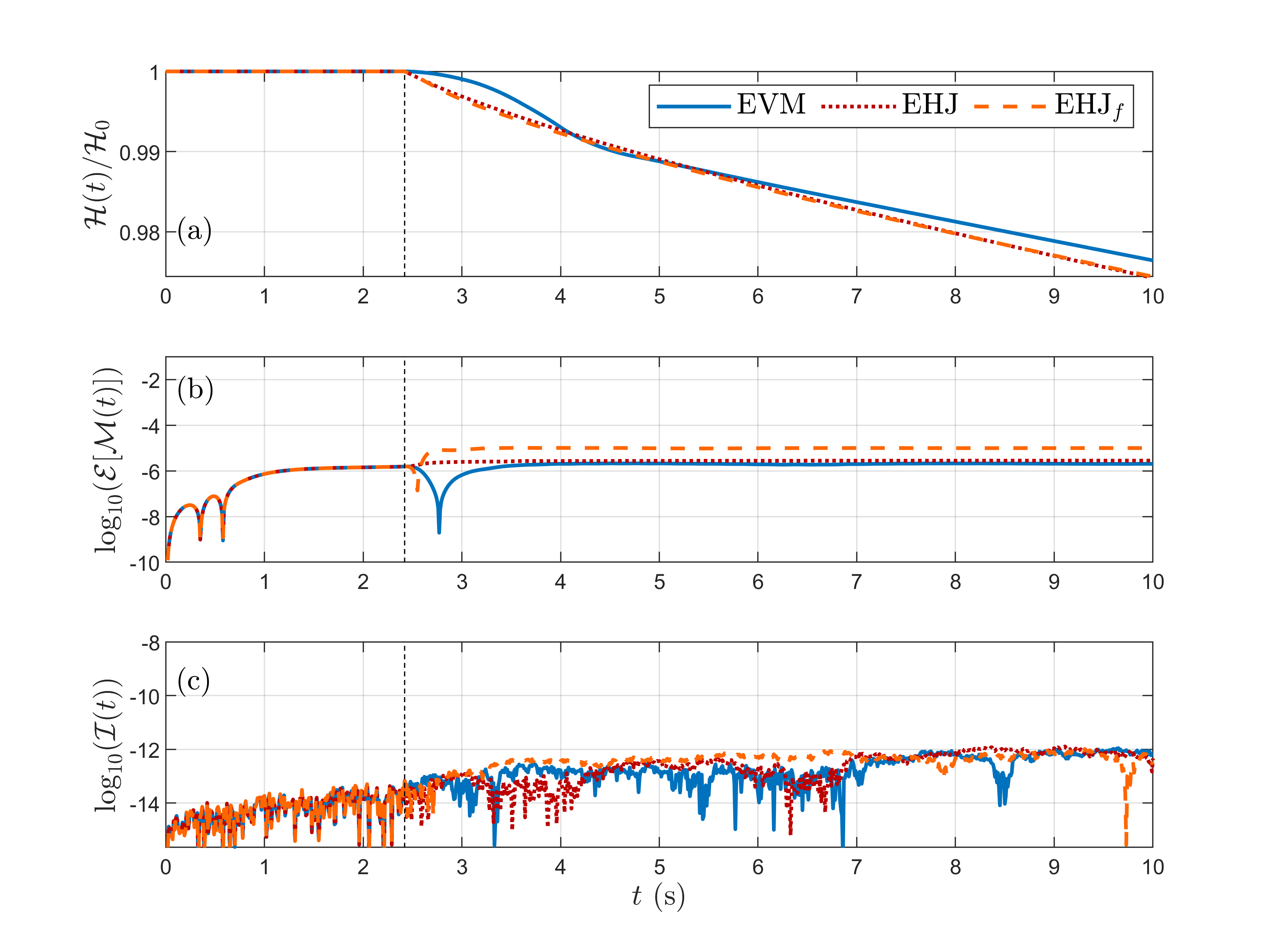}
		\caption{Evolution of (a) the Hamiltonian, (b) the logarithm of the relative error on the mass, and (c) the logarithm of the horizontal momentum for the simulation of two counter-propagating breaking DSWs. The vertical dashed line indicates the initiation of breaking.}
		\label{Fig_Impulse_new}
	\end{figure}

	 Figure \ref{Fig_Impulse_new} shows the ratio $\mathcal{H}(t)/\mathcal{H}_0$ (where $\mathcal{H}_0$ is the initial total energy), the logarithm of the relative error of the mass, $\mathcal{E}\left[\mathcal{M}(t)\right]=\left|\mathcal{M}(t)/\mathcal{M}_0-1\right|$ and the logarithm of the momentum $\mathcal{I}(t)$. Both methods dissipate energy and conserve momentum. Mass is also well conserved in the simulations. The EHJ and EHJ$_f$ methods dissipate energy at the same constant rate.  This is expected since, in this method, the extent of the breaking region does not affect the amount of dissipated energy. In the EVM approach, the dissipation rate increases gradually during a transition period after which it remains constant. 
	
	\subsubsection{Dam-break experiment of \normalfont\cite{Carmo93}}
\begin{figure}
				\centering
				\includegraphics[scale = 0.6]{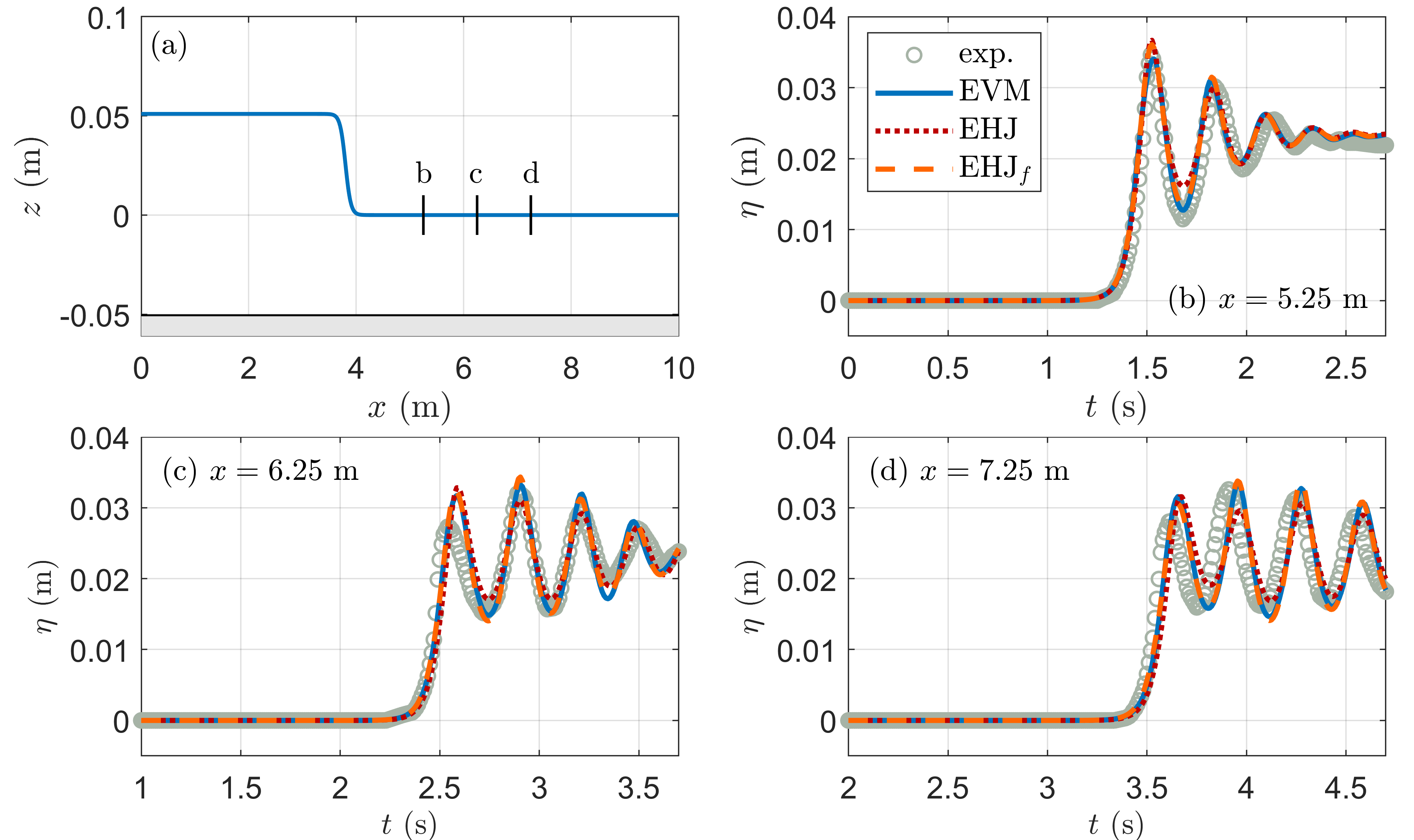}
				\caption{(a) Initial configuration for the experiment of \cite{Carmo93}. Vertical lines mark the locations of the measuring gauges. (b,c,d) Comparison of computed free surface elevation with experimental measurements.}
				\label{Fig_CARMO}
			\end{figure}

		Next, the dam-break experiment of \cite{Carmo93} is simulated. In a rectangular wave tank, two different volumes of water of heights $H_2=0.099$ m and $H_1=0.051$ m are separated by a gate or dam. Upon the sudden removal of the dam, a dispersive shock wave propagates downstream. As shown in \cite{CFP2019}, both wave breaking and strong dispersive effects play an important role in this experiment. For the simulation, the removal of the dam is modeled by considering an initial free-surface elevation of the form
		\begin{align}
		\eta(x,0) = \frac{1}{2}(H_2-H_1)\left(1-\tanh\left(\frac{x_0-x}{s}\right)\right),
		\end{align}
		where $x_0=3.8$ m is the location of the gate, and $s=0.1$ m (see Figure \ref{Fig_CARMO} (a)).  Computations are performed by using $\delta x = 0.01$ m, $\delta t = 0.01$ s and $N_{\text{tot}}=5$ vertical functions. Wave breaking is initiated with $\gamma_{\text{I}}=0.6$. Note that the value $\gamma_{\text{F}}=0.3$ used for the termination criterion was not reached during the duration of the simulation.
		
		Figure \ref{Fig_CARMO} compares the computed free-surface elevation with the experimental measurements at three locations. Note that a time shift of 0.2 s was needed to match the results at the first station, and this time shift is therefore applied at all stations. As noted in \citet{CARMO2018}, this lag is may be explained by the physical difference between the initial condition used in the simulations and the removal of the dam in the experiment of \cite{Carmo93}. All computations agree well with the experiments at the first station (Figure \ref{Fig_CARMO}(b)) but overestimate the height of the front wave at the next two stations (Figures \ref{Fig_CARMO}(c) and (d)). At these two stations, EHJ underestimates the wave heights of the waves following the breaking front, while EVM and EHJ$_f$ perform better. At the last station, a small phase shift is observed for all methods (Figure \ref{Fig_CARMO}(d)), which likely can be attributed to the difference between the initialization of the experiment and the simulation. 
		
	\subsection{Breaking of shoaling regular waves on a sloping beach \normalfont\citep{TK94}}\label{SubsecTK}
\begin{figure}
		\centering
		\includegraphics[scale=0.6]{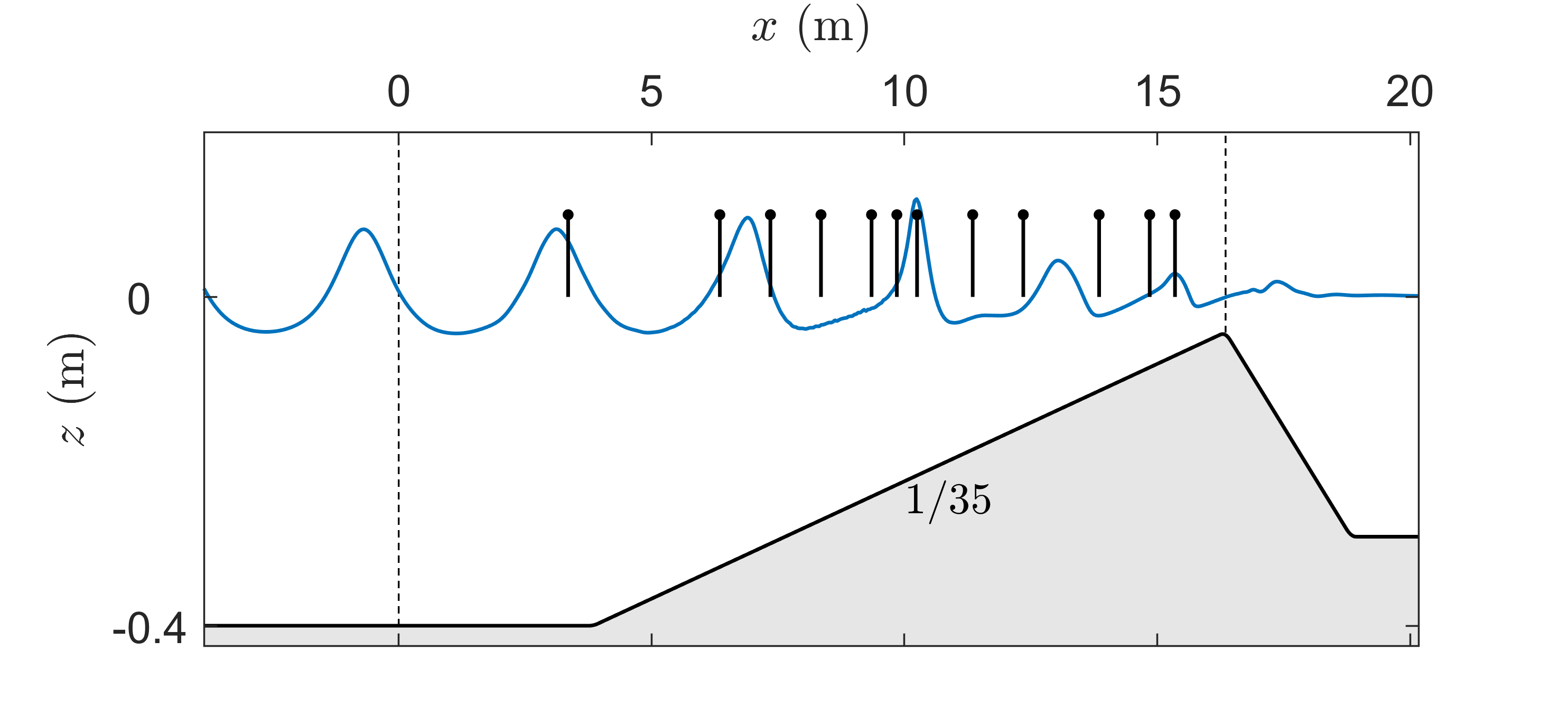}
		\caption{Bathymetry used in the simulations to represent the experiment of \cite{TK94}. Vertical lines correspond to the wave gauge locations. Sponge layers are indicated by the vertical dashed lines.}
		\label{Fig_TK_config}
	\end{figure}

	In this section, the spilling breaking experiment of \cite{TK94} is considered. Incident waves of wave height $H=0.125$ m and  period $T=2.0$ s (wavelength $L=3.85$ m) are generated in constant depth $h_0=0.4$ m and propagate towards a plane beach of slope 1/35. Non-synchronised measurements of the free-surface elevation, as well as the corresponding mean crest, trough, and water levels are available at 21 locations along the wave tank. Due to the absence of a moving shoreline in the current numerical model, the original geometry is modified in the simulation: after reaching a minimum depth $h=0.0434$ m, an absorbing sponge layer is applied, and in this zone the bathymetry deepens again with a slope 1/10 until $h=0.2915$ m (Figure \ref{Fig_TK_config}). Simulations are performed using $N_{\text{tot}}=4$ vertical functions, a spatial grid with $\delta x = L/150=0.0257$ m, and a time step $\delta t = T/200=0.01$ s. Breaking is initiated with $\gamma_{\text{I}}=0.6$ and terminated with $\gamma_{\text{F}}=0.1$. Note that breaking terminates inside the sponge layer. The variant EHJ$_f$ of EHJ, where only the wave front is considered as the breaking region, rapidly becomes unstable, and the simulations could not be completed.  

\begin{figure}
		\centering
		\includegraphics[width=\textwidth]{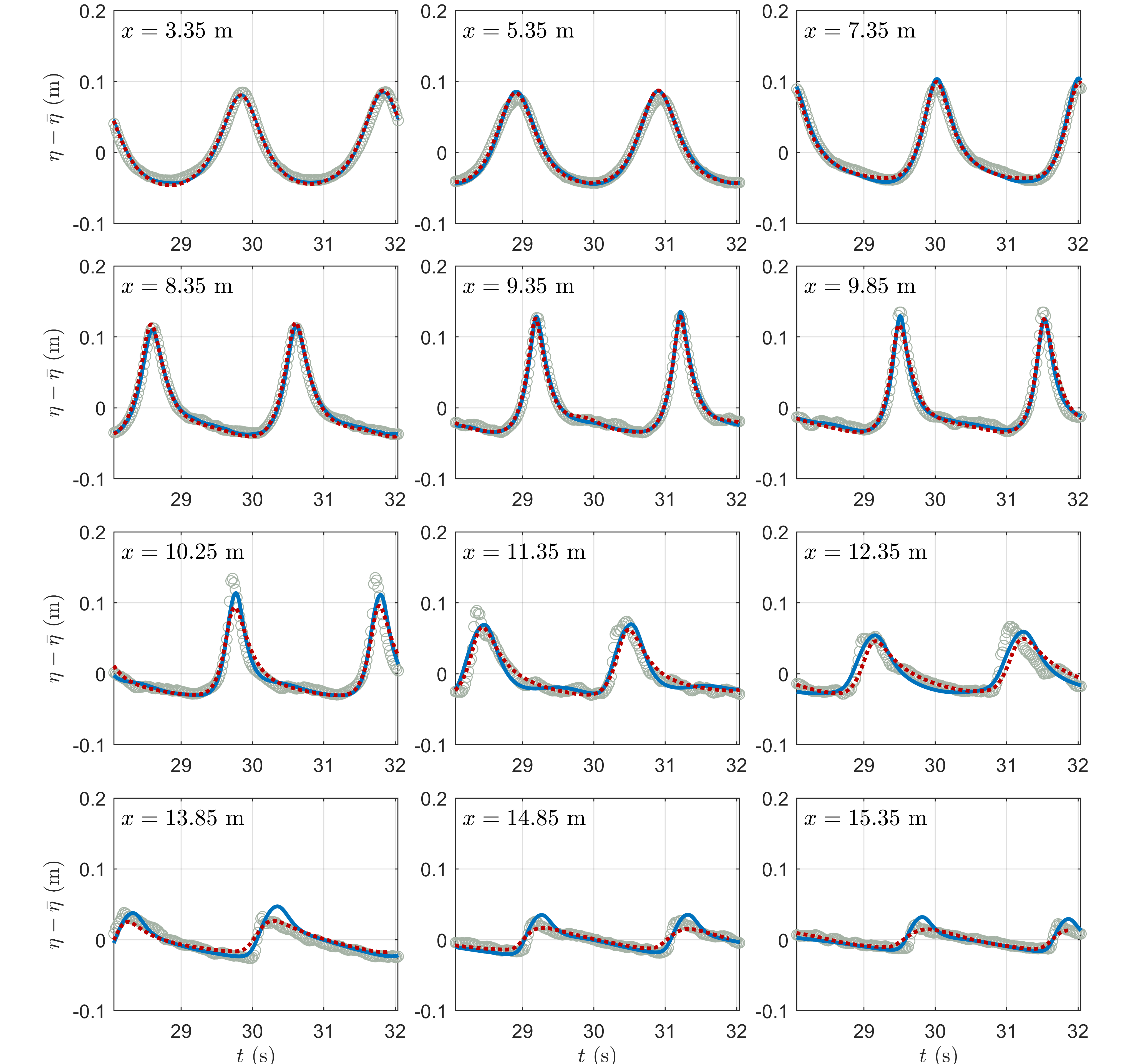}
		\caption{Comparison of experimental data (circles) and the simulated EVM (plain lines) and EHJ (dotted lines) time series of the free surface elevation at the locations shown in Figure \ref{Fig_TK_config} for the case of shoaling and spilling breaking of regular waves propagating over  a mildly sloping beach \citep{TK94}. Breaking begins around $x=10.25$ m}.
		\label{Fig_TK_comp}
	\end{figure}

	In Figure \ref{Fig_TK_comp}, the computed free-surface elevation is compared with the experimental measurements at 12 locations spanning the shoaling and surf zones (Figure \ref{Fig_TK_config}). The computed and measured free-surface elevation profiles agree well up to $x=9.85$ m. In the vicinity of the breaking point, around $x=10.25$ m, the EVM slightly underestimates the wave height, while the EHJ method significantly underestimates the wave height. Some discrepancies in the wave shape are also noticeable at the last four stations for both wave breaking approaches.
	
\begin{figure}
		\centering
		\includegraphics[width=0.8\textwidth]{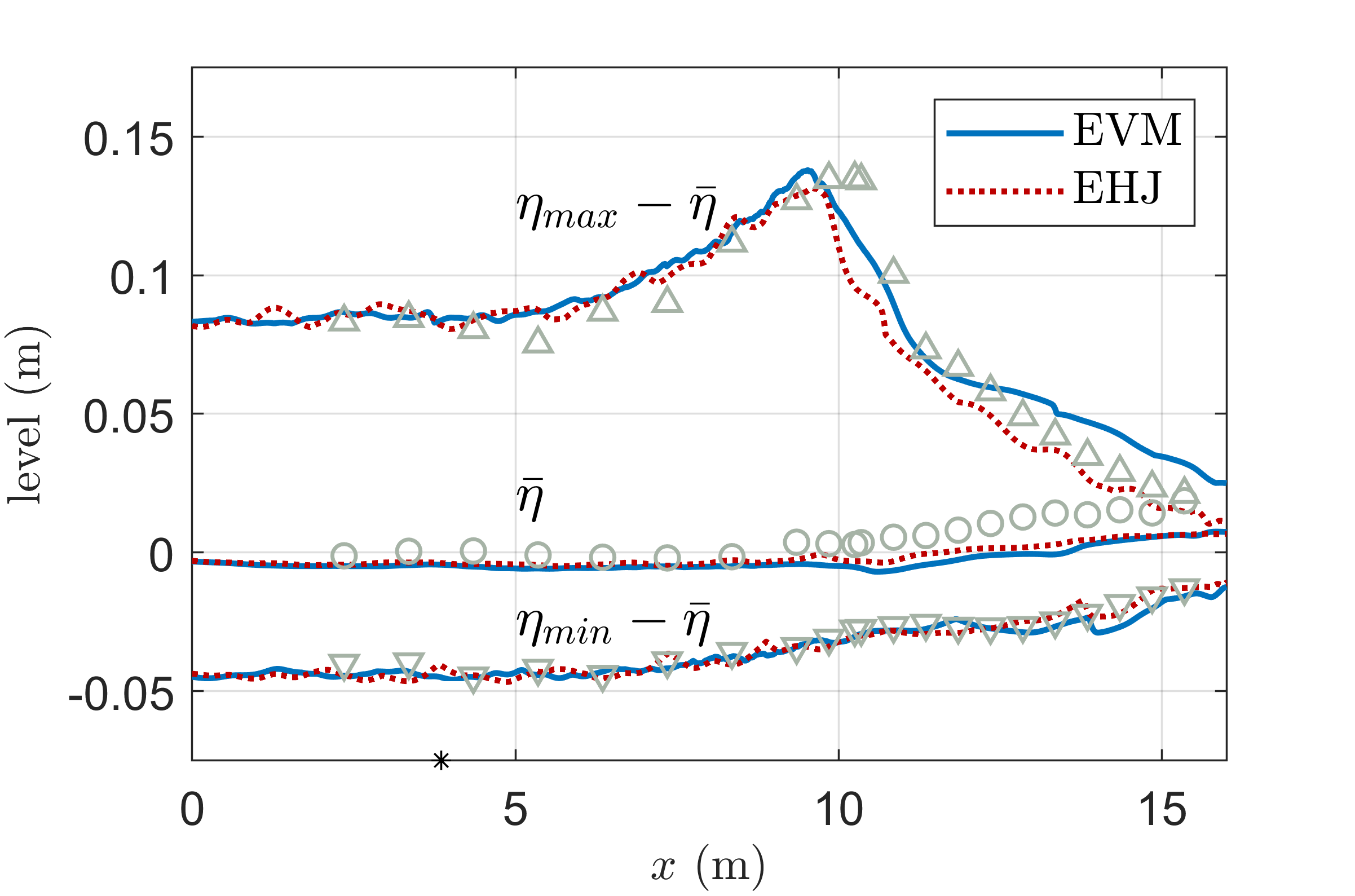}
		\caption{Simulated spatial variation of the Mean Water Level (MWL) $\bar{\eta}$,  maximum elevation relative to MWL $\eta_{max}-\bar{\eta}$, and minimum elevation relative to MWL  $\eta_{min}-\bar{\eta}$, for the experiment of \cite{TK94}. The asterisk indicates the toe of the beach. Numerical results are averaged over $10T$ sec.}
		\label{Fig_TK_levels}
	\end{figure}

	The maximum and minimum of the spatial variations of the time-averaged surface elevation levels are reproduced well (Figure \ref{Fig_TK_levels}). Breaking in the simulations occurs slightly earlier ($x=9.7$ m) than observed in the experiments ($x=10.25$ m). This likely explains why the simulations underestimate the wave heights at $x=10.25$ m in Figure \ref{Fig_TK_comp}. Also at this location, the more significant underestimation of the wave height by the EHJ method is explained by the abrupt application of wave breaking energy dissipation, while in the EVM it is applied smoothly in time. Concerning the prediction of set-up, both simulations agree well with the laboratory data before the breaking point, but underestimate the increase in the mean water level in the surf zone (Figure \ref{Fig_TK_levels}). This trend is also visible in other modeling approaches \citep{CRB10,TBMCL12}. 

\begin{figure}
	\centering
	\includegraphics[width=0.8\textwidth]{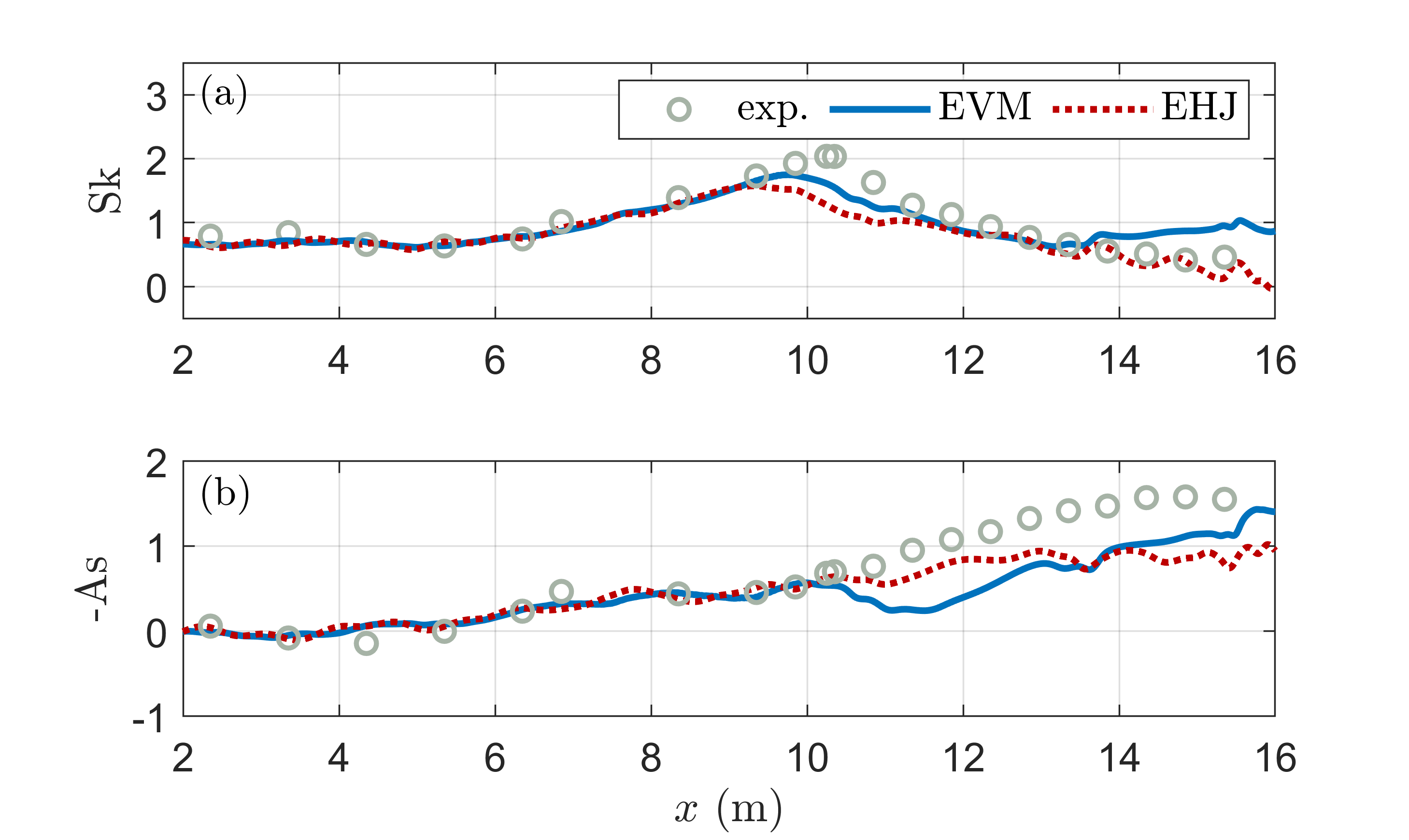}
	\caption{Spatial variation of the (a) wave skewness Sk and (b) wave asymmetry As for the experiment of \cite{TK94}. Numerical results averaged over $10T$ sec.}
	\label{Fig_TK_nlp}
\end{figure}

	To further assess the ability of the proposed models to reproduce accurately the nonlinear wave shape, the skewness Sk and asymmetry As of the surface elevation are calculated following \citet{Kennedyetal2000}:
	\refstepcounter{equation}
	 $$
	\text{Sk} = \frac{\left<(\eta-\bar{\eta})^3\right>}{\left<(\eta-\bar{\eta})^2\right>^{3/2}} ,\qquad
	\text{As} = \frac{\left<\left[\,\mathbb{H}(\eta-\bar{\eta})\,\right]^3\right>}{\left<(\eta-\bar{\eta})^2\right>^{3/2}},\eqno{(\theequation{\mathit{a},\mathit{b}})}\label{eq:SkAS}
	$$  
	where $\left<\,\right>$ is the time-averaging
	operator, $\bar{\eta}=\left<\eta\right>$ and $\mathbb{H}$ denotes the Hilbert transform. Figure \ref{Fig_TK_nlp} shows a comparison of the measured and computed spatial variation of Sk and -As. For the EVM, the skewness Sk is  reproduced accurately in the shoaling region and in a portion of the surf zone spanning from $x=11.35$ to 13.85 m. Differences appear in a small region after the breaking point and in the last 2-3 meters of the inner surf zone. Concerning the wave asymmetry As, the EVM simulation results agree well with the experimental measurements until the breaking point, after which it is underestimated. Similar trends are also present in computations using a BT model with Kennedy's eddy viscosity approach \cite[Figures 2,4]{CRB10} and in computations using a GN model with the Hybrid approach \cite[Figure 7]{TBMCL12}. Using the EHJ approach, the wave skewness is reproduced well except for a small region around the breaking point. The simulated wave asymmetry also agrees well with the experimental measurements in the shoaling region but presents some differences after breaking begins. In comparison to the EVM, the EHJ method generally predicts smaller values of Sk, and first smaller and then ($x\geq 13.5 $ m) larger values of -As in the region after the breaking point.
	
	On the basis of the above results, both methods predict accurately the wave height evolution during the shoaling/breaking process in comparison with BTM approaches (see e.g \citet{TBMCL12}, Figure 7 (a)) or the Reynold Averaged Navier-Stokes approach (see \citet{DKSM16}, Figure 2(a)). Differences in the wave shape in the surf zone do exist and, as mentioned earlier, this drawback is also present in the original eddy viscosity approach (see the discussion in \cite{CRB10}) and the Hybrid approach applied in BTMs (see e.g. \citet{TBMCL12}, Figure 7 (c)).
It is observed that both EVM and EHJ methods induce a dissipation rate that is not optimal for the spilling breaking case. Controlling the dissipation rate by considering breaking regions of arbitrary length may be one possible option to investigate. Note, however, this would introduce 
 at least one more free parameter (breaking region length) in the resulting breaking model.

	\subsection{Regular waves breaking over a bar \normalfont\citep{BB93}}\label{SubsecBB}
\begin{figure}
		\centering
		\includegraphics[scale=0.6]{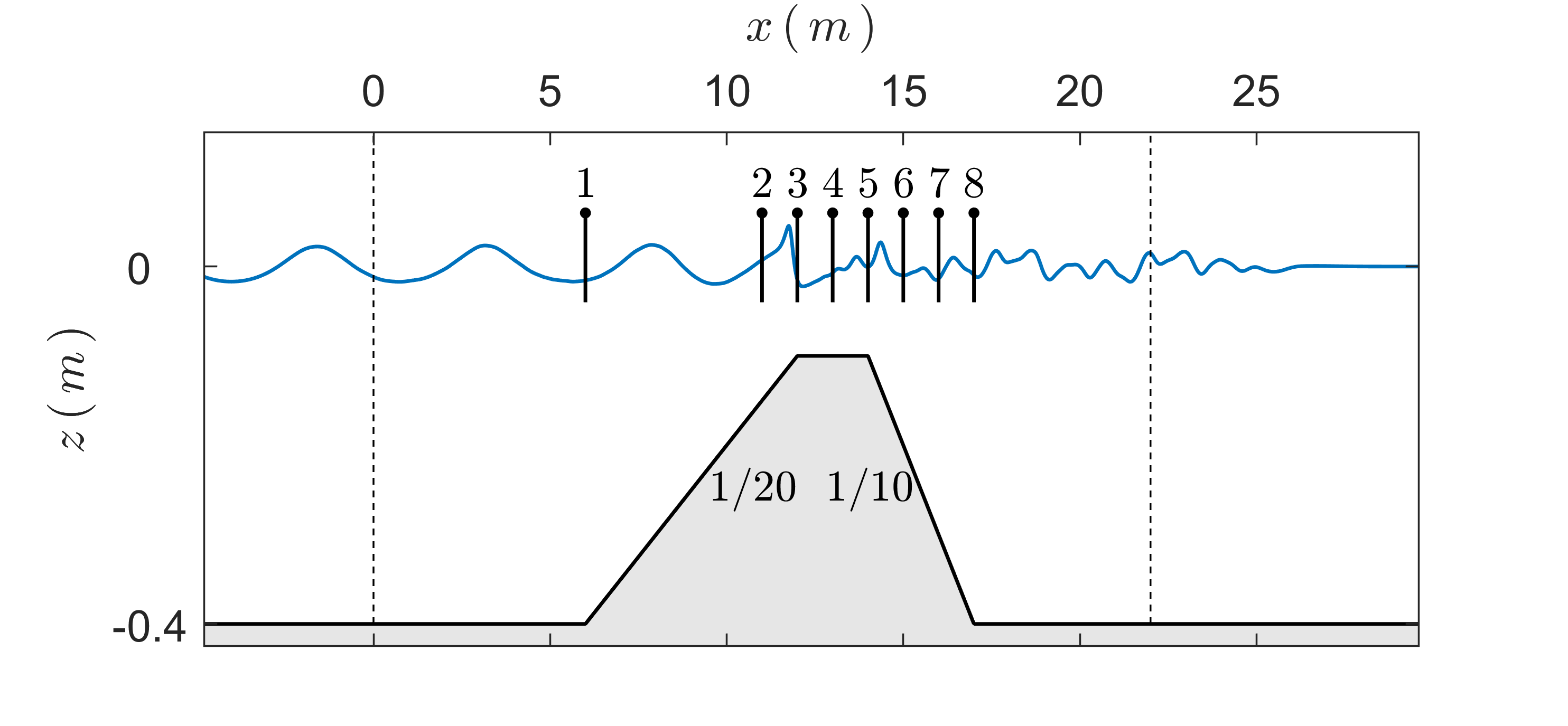}
		\caption{Bathymetry in the experiment of \cite{BB93}. Vertical lines indicate the locations of the wave gauges. Vertical dashed lines indicate the extent of the sponge layers.}
		\label{Fig_BB_config}
	\end{figure}

	\citet{BB93} investigated the transformation of periodic waves propagating over a submerged trapezoidal bar (Figure \ref{Fig_BB_config}). In the experiments, the incident wave train shoals along the front face of the bar, and bound harmonics are amplified. Plunging breaking occurs over the top of the bar, and then higher harmonics are released on the lee side of the bar, creating a strongly dispersive wave form. The case simulated here corresponds to a long wave generated in a constant depth $h_0=0.4$ m with period $T=2.5$ s ($f=0.4$ Hz, wavelength $L=4.8$ m) and wave height $H=0.054$ m. For the simulations, the spatio-temporal discretization is $\delta x = L/180=0.0267$ m, $\delta t = T/180=0.014$ s, and $N_{\text{tot}}=6$ vertical functions are used to compute the DtN operator. Breaking is initiated with $\gamma_{\text{I}}=0.3$ and terminated with $\gamma_{\text{F}}=0.1$.
	
\begin{figure}
		\centering
		\includegraphics[width=\textwidth]{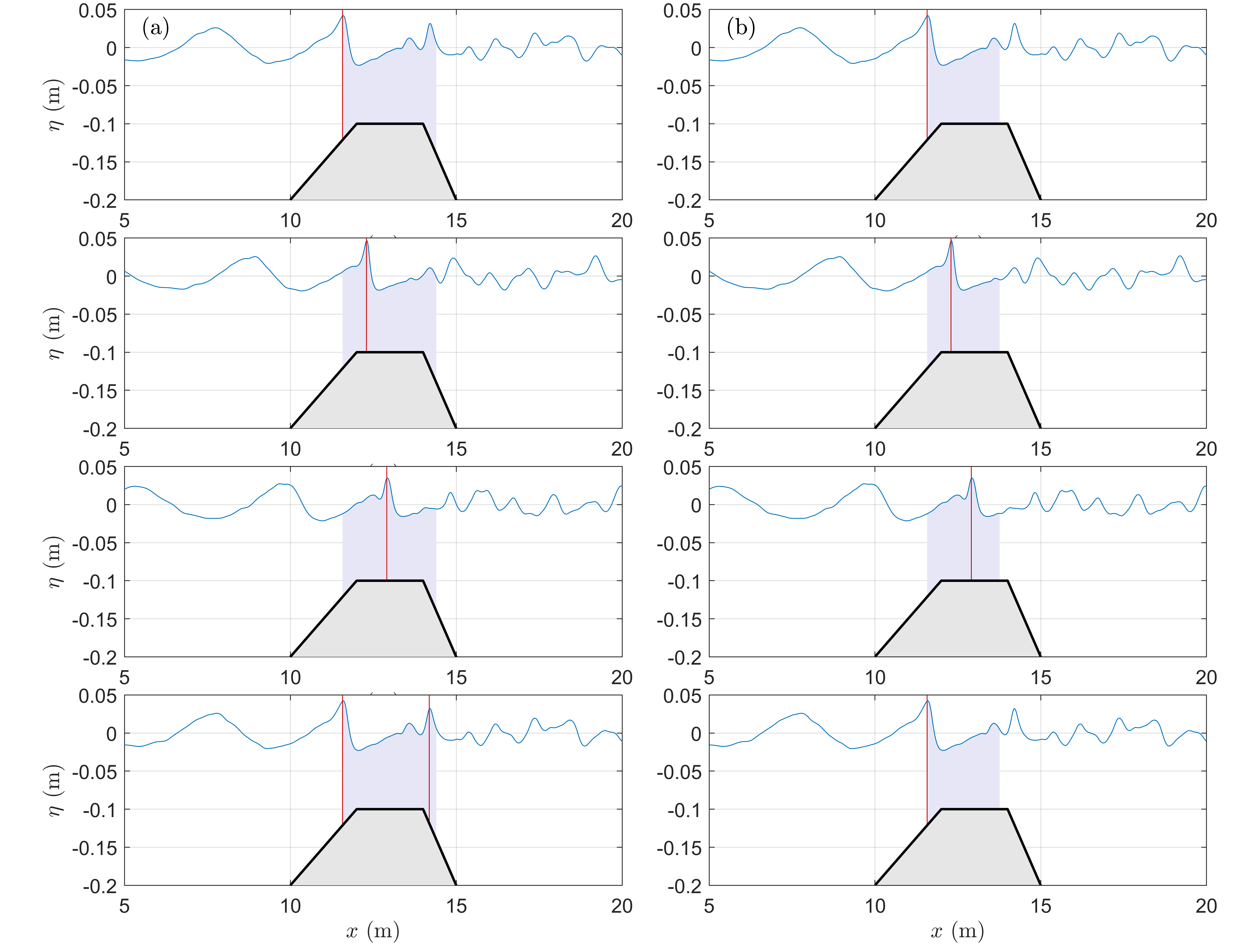}
		\caption{Snapshots of the simulated free-surface elevation for the experiment of \cite{BB93} using EVM (column (a)) and EHJ (column (b)). Vertical lines correspond to the crests of the breaking waves. The shaded area indicates the spatial extent of wave breaking in each simulation.}
		\label{Fig_BB_snap}
	\end{figure}

	Figure \ref{Fig_BB_snap} shows snapshots of the computed free surface elevation using the EVM and EHJ. Incident waves start breaking before the top of the bar, at $x=11.57$ m. In the case of the EVM, the waves continue breaking until reaching $x=14.4$ m, after passing the top edge of the bar. During this phase, the next incoming wave starts breaking, and two breaking fronts exist simultaneously. This is also observed when using the EHJ$_f$ (not shown here) and agrees with the simulation results shown by \cite{KDS14}. However, when using the EHJ, wave breaking stops sooner, at $x=13.76$ m, and only a single breaking wave exists at any moment in time during the simulation. 

\begin{figure}
		\centering
		\includegraphics[width=\textwidth]{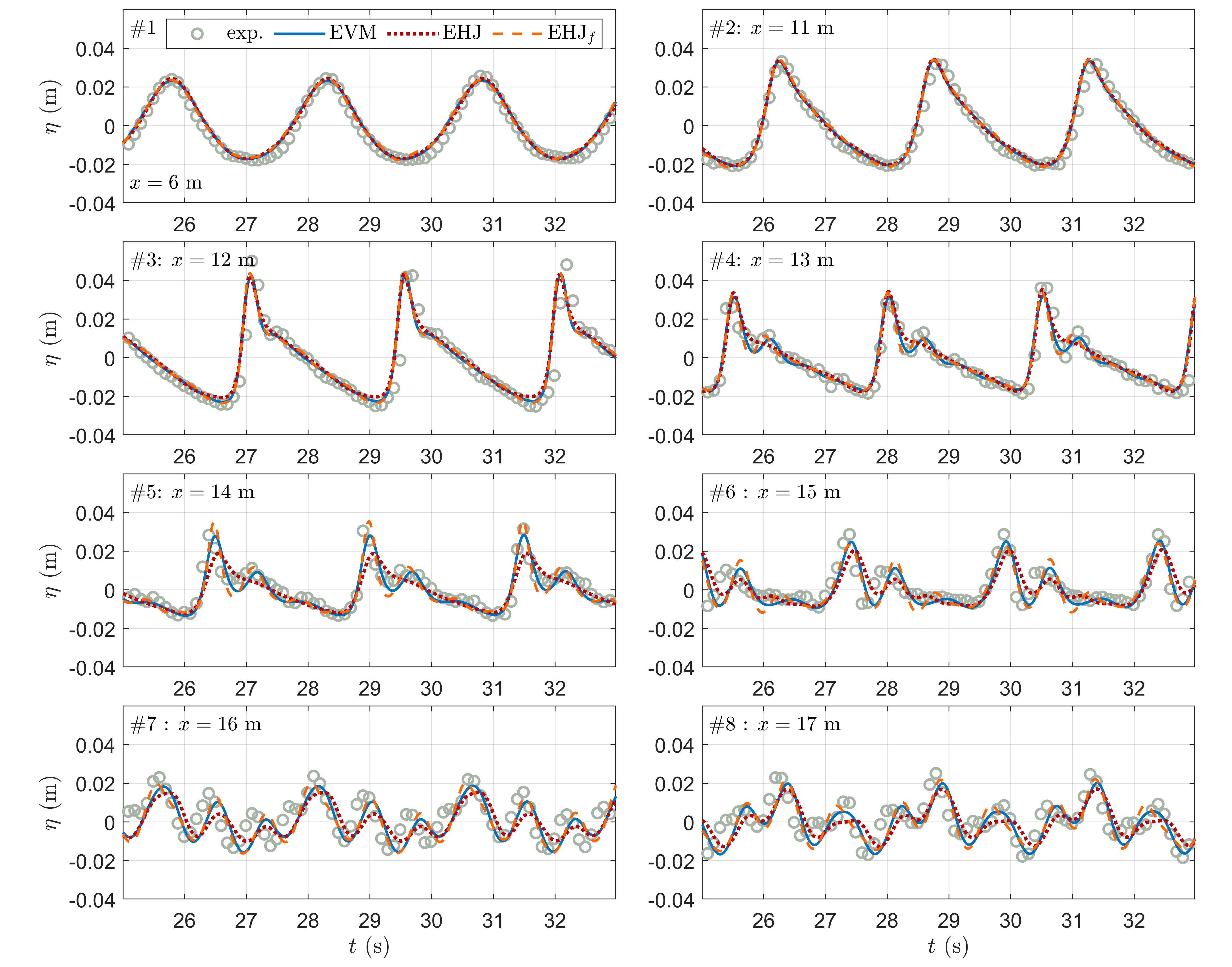}
		\caption{Comparison of experimental data (circles) and the EVM (plain lines), EHJ (dotted lines) and EHJ$_f$ (dashed lines) simulation results. Free-surface elevation time series at locations \#1-\#8 for the transformation and plunging breaking of regular waves propagating over a submerged bar \citep{BB93}.}
		\label{Fig_BB_comp}
	\end{figure}

	The simulated time series of the free-surface elevation at the eight wave gauge locations are compared with the experimental measurements in Figure \ref{Fig_BB_comp}. Good agreement with the experimental
	measurements is obtained for all three simulation methods. Differences are visible only at stations \#4 and \#6, where EVM and EHJ$_f$ capture the wave transformation processes slightly more accurately than EHJ. 
	
\begin{figure}
	\centering
	\includegraphics[scale=0.6, ]{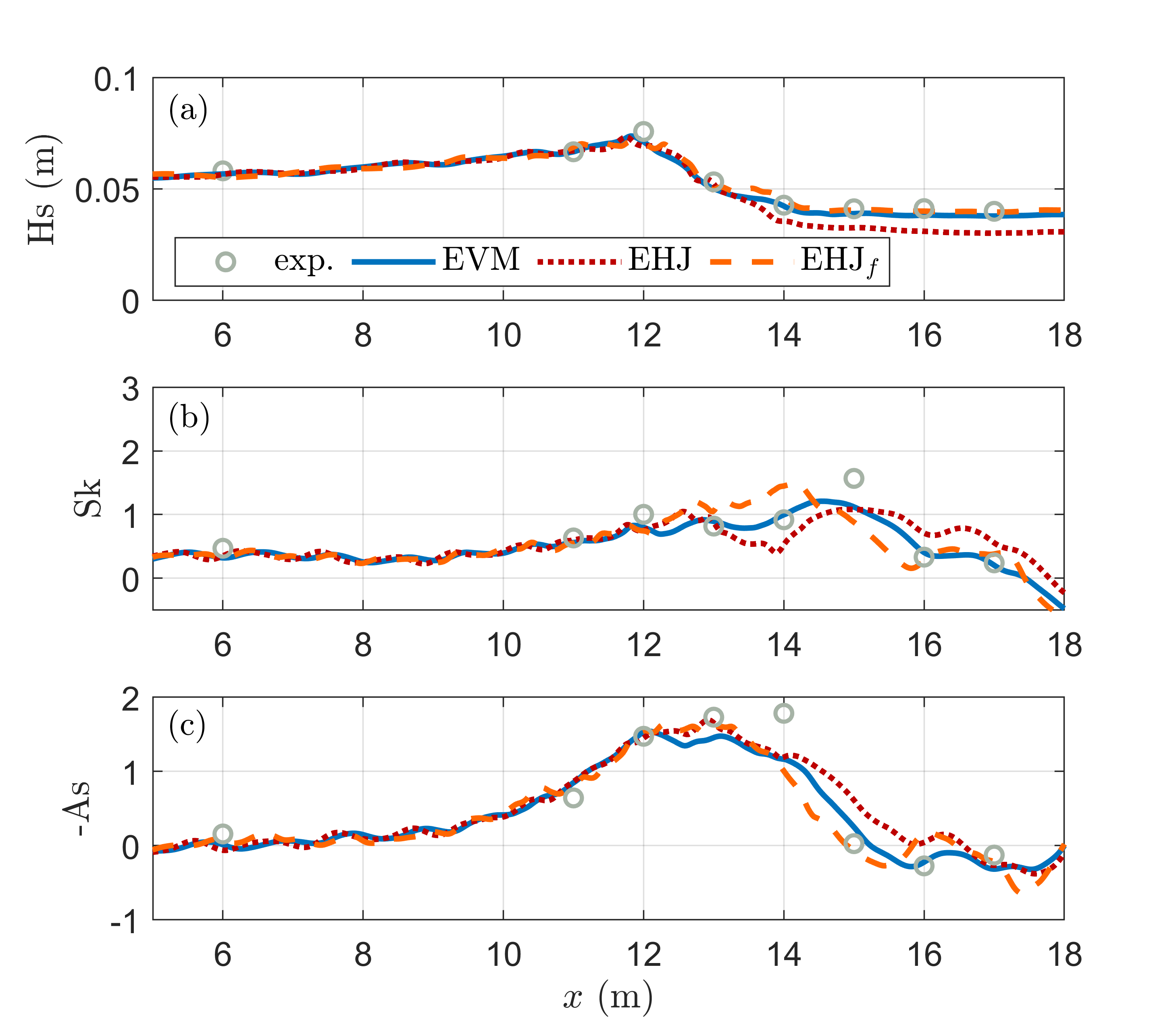}
	\caption{Spatial evolution of the (a) significant wave height Hs, (b) wave skewness Sk, and (c) wave asymmetry As, calculated over $15T$ sec, for the experimental measurements (circles), EVM (plain lines), EHJ (dotted lines), and EHJ$_f$ (dashed lines) of the experiment of \cite{BB93}.}
	\label{Fig_BB_nlp}
\end{figure}

	Figure \ref{Fig_BB_nlp} shows the spatial variation of the significant wave height Hs (calculated as four times the standard deviation of the surface elevation), skewness Sk and asymmetry As (Eq. \eqref{eq:SkAS}). These nonlinear wave characteristics are reproduced well when using the EVM. The EHJ underestimates Hs after breaking and provides a fair prediction of Sk and As. The EHJ$_f$ reproduces well  Hs and As but shows some differences in Sk after breaking.
	
\begin{figure}
	\centering
	\includegraphics[scale=0.9 ]{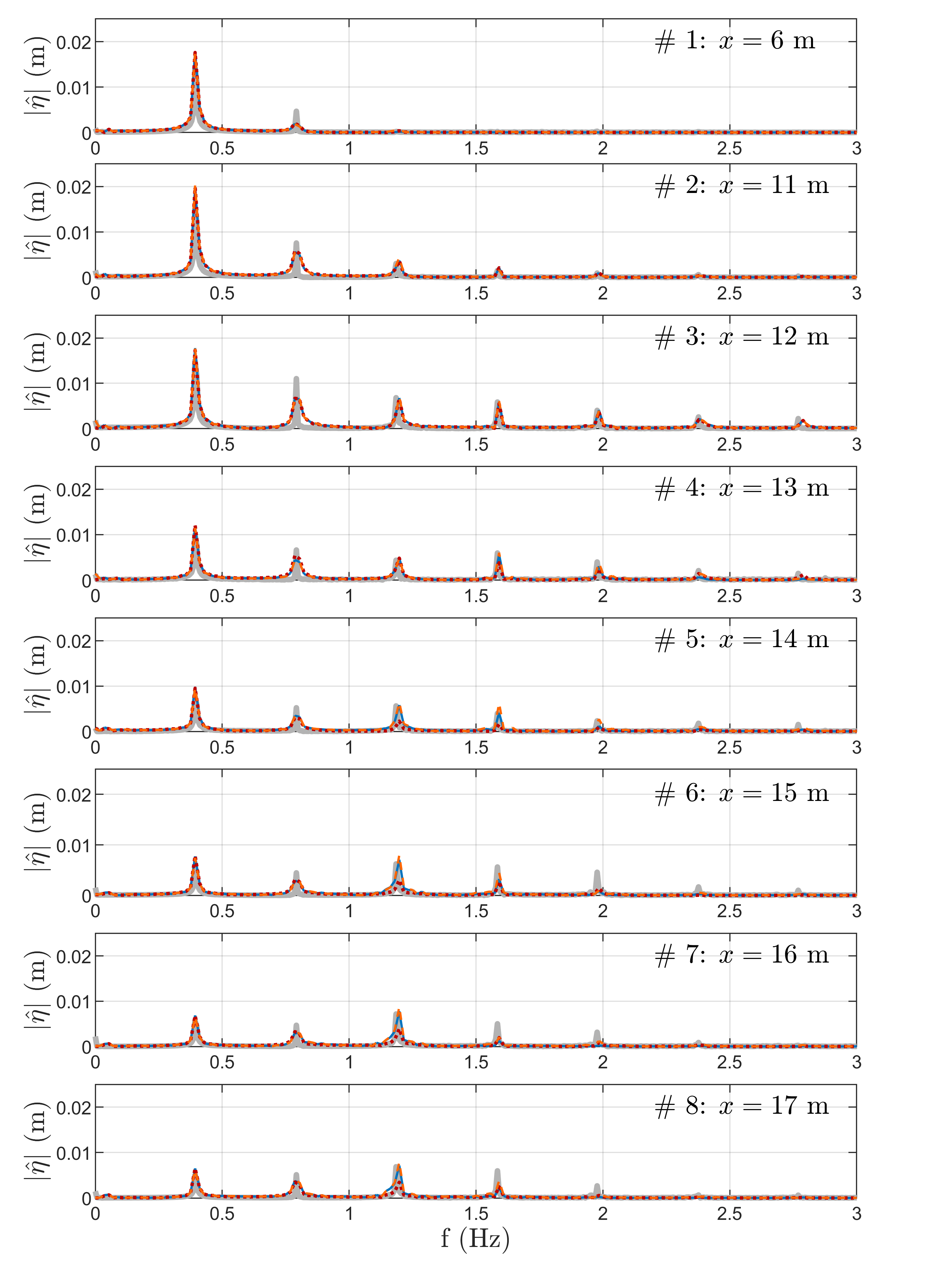}
	\caption{Comparison of the wave spectrum between the experimental data (grey line), EVM (plain lines), EHJ (dotted lines), and EHJ$_f$ (dashed lines) for the experiment of \cite{BB93}.}
	\label{Fig_BB_comp_FA}
\end{figure}

Next, a spectral analysis of the computed and experimental time-series of the free-surface elevation  is performed in order to assess the energy transfers between harmonics and the wave decomposition process (Figure \ref{Fig_BB_comp_FA}). Before the breaking point (wave gauge \#2, $x=11$ m) and for the next two stations, all three methods agree well with each other and with the experiment, capturing the energy transfer processes to higher harmonics, with the exception of a small underestimation of the second harmonic amplitude. Starting from wave gauge \#5, the EVM and EHJ$_f$ approaches produce similar results, accurately reproducing the wave decomposition process, while the EHJ approach underestimates the amplitude of the third harmonic.
		
	Taking into account the complexity of the physical processes (e.g. violent air-water mixing during post breaking evolution), the EVM and EHJ$_f$ reproduce well the free-surface effects. Using the EVM, differences are obtained for the wave skewness at wave gauge \#6 and for the wave asymmetry at wave gauge \#5. Differences for the EHJ$_f$ approach are obtained for wave skewness at wave gauges \#5 and \#6 and for the asymmetry at wave gauge \#5. Finally, it is important to note that, although a small phase error is obtained at the last two wave gauges where dispersive effects are most important, both methods reproduce accurately the wave skewness and asymmetry. 
	

	\section{Discussion and Conclusions}\label{Sec:Conclusions}
	In this paper, two different methods are presented for simulating the effects of wave breaking in the framework of the Hamiltonian formulation of free-surface potential flow. The first, called the EVM, is a variant of the eddy viscosity model proposed for BT models by \cite{Kennedyetal2000}. The term introduced in the FSBC is computed by solving a differential equation derived by requiring that momentum is conserved as the waves break over a flat bottom. The second method, EHJ, is the implementation of the dissipative term proposed by \citet{GG2001} and \citet{GG2019}. This term is constructed such that it produces negative work proportional to the work done by a hydraulic jump with wave characteristics similar to those of the breaking wave. For both methods, wave breaking events are determined by simple initiation and termination criteria as a function of the velocity of the free-surface elevation. The numerical scheme is based on the Hamiltonian Coupled-Mode Theory, which is implemented with a fourth-order spatial discretization and an explicit fourth-order Runge-Kutta time-stepping algorithm. 
	
	The proposed breaking models were applied to several test cases. The first test case verified numerically that both methods dissipate energy while conserving momentum for the propagation over a flat bottom of breaking DSWs. A laboratory dam-break problem was also successively reproduced by all three methods, with the EVM producing the best results. The overall performance of the EVM was good for all examined cases, giving particularly accurate results in the case of a plunging wave breaking over a submerged trapezoidal bar. Fair agreement with the same experiment is also obtained with the EHJ, even though this approach was originally proposed for spilling breakers on milder slopes. Here, it was also shown that restricting the breaking region of the EHJ to the front face of the breaking wave (EHJ$_f$ variant) improves results for dispersive shock waves and plunging breakers propagating over a barred bathymetry. In the case of spilling breaking waves propagating over a mildly sloping beach, the EVM and EHJ methods reproduced well the maximum and minimum levels of the free-surface elevation. An underestimation of the  increase in the mean water level after the breaking point (set-up) is obtained by both methods. This could be explained partially by the inability of the present numerical scheme to take into account a moving shoreline and by the use of a modified bathymetry and a sponge layer at the end of the wave tank. 
	More generally, the application of energy dissipation as a pressure term into the Bernoulli FSBC may not introduce the optimal radiation stress changes in the case of spilling breakers.
	 These issues deserve further investigation and should be the focus of future work.
	
	To conclude, the proposed methods demonstrate promising results in water wave problems where strong nonlinearity, dispersion and wave dissipation are present. They are quite general in the sense that they can be implemented in other wave models provided that they use the free-surface potential as one of the evolving variables. Future work includes extending the present one-dimensional (1DH) approach to two horizontal dimensions (2DH). Although the performances of the two presented methods are comparable in 1DH, the extension of EVM to 2DH seems more straightforward because it does not depend on the instantaneous geometric characteristics of breaking waves. Additional future work includes the incorporation of bottom friction effects, the treatment of a moving shoreline (run-up), and the application of the models to realistic coastal applications.

	\section*{Acknowledgements}
	The authors thank Prof. do Carmo (University of Coimbra, Portugal) for providing the laboratory data for the dam-break problem, Prof. J. T. Kirby (University of Delaware, USA) for providing the laboratory data of regular waves breaking on a sloping beach and Prof. S. Beji (Istanbul Technical University, Turkey) for providing the laboratory data for waves breaking over a submerged bar. This work benefited from France Energies Marines and French State financing managed by the National Research Agency under the Investments for the Future programme bearing the reference ANR- 10-IED-0006-18. The work of B. Simon has been carried out in the framework of the Labex MEC and received funding from Excellence Initiative of Aix-Marseille University - A*MIDEX, a French ``Investissements d'Avenir" programme.
	\appendix

	\section{Proof of Eq. \eqref{eq:IdE}}\label{App:dE_dxH}
	Differentiation of the kinetic energy term of $E$, $K = 1/2\,\psi\, \mathcal{G}[\eta]\psi$, with respect to $x$ gives
	 \begin{align}\label{eq:2Kx}
	 2\partial_{x}K = \partial_{x}\psi\, \mathcal{G}[\eta]\psi + \psi\, \partial_{x}\left(\mathcal{G}[\eta]\psi\right).
	 \end{align}
	 The expression for the $x$-derivative of the DtN operator derived in \citet[Corollary 3.26]{Lannesb} can be written as
	 \begin{align}\label{eq:Gx}
	 \partial_{x}(\mathcal{G}\psi) = \mathcal{G}(\partial_x\psi-\mathcal{W}\psi\,\partial_{x}\eta)-\partial_x\big[(\partial_x\psi-\mathcal{W}\psi)\,\partial_{x}\eta\big],
	 \end{align}
	  where the simplified notation $\mathcal{G}[\eta]\psi\equiv\mathcal{G}\psi$ is introduced together with the vertical derivative of the velocity potential on the free surface,
	 \begin{align}\label{eq:Wpsi}
	 \mathcal{W}\psi\equiv\mathcal{W}[\eta]\psi = \frac{\mathcal{G}\psi+\partial_{x}\eta\partial_{x}\psi}{1+(\partial_{x}\eta)^2}=\left[\partial_z\Phi\right]_{z=\eta}.
	 \end{align}
	 Using Eq. \eqref{eq:Gx}, the right hand side of Eq. \eqref{eq:2Kx} takes the form
	 \begin{align}\label{eq:2Kxrhs}
	 \partial_{x}\psi\,\mathcal{G}\psi + \psi\,\mathcal{G}(\partial_{x}\psi) - \psi\,\mathcal{G}(\mathcal{W}\psi\partial_{x}\eta)-\psi\partial_{x}\big[\big(\partial_{x}\psi-\mathcal{W}\psi\,\partial_{x}\eta\big)\partial_{x}\eta\big].
	 \end{align}
	 Integrating Eq. \eqref{eq:2Kx}, we obtain 
	 \begin{align}
	 \int 2\partial_{x}K dx = \int\left\{2\partial_{x}\psi\,\mathcal{G}\psi - \mathcal{G}\psi\,\mathcal{W}\psi\,\partial_{x}\eta + \partial_{x}\psi\big(\partial_{x}\psi - \mathcal{W}\psi\,\partial_{x}\eta\big)\partial_{x}\eta \right\}dx,
	 \end{align}
	 where the self-adjointness of $\mathcal{G}$ has been used to derive the second and third terms in Eq. \eqref{eq:2Kxrhs} and the vanishing at infinity condition to derive the fourth term. By simply rearranging the above equation, it takes the form
	 \begin{align}\label{eq:int2Kx}
	 \int 2\partial_{x}K dx = \int\left\{2\partial_{x}\psi\mathcal{G}\psi + \partial_{x}\eta\Big[(\partial_{x}\psi)^2 - \left(\partial_{x}\eta\partial_{x}\psi+\mathcal{G}\psi\right)\mathcal{W}\psi\Big]\right\}dx.
	 \end{align}
	 Differentiation of the potential energy term of $E$, $V=1/2g\eta^2$, gives
	 \begin{align}\label{eq:int2Vx}
	 \int 2\partial_{x}V dx = 2 g\int \eta\partial_{x}\eta dx.
	 \end{align}
	 Combining Eqs. \eqref{eq:int2Kx} and \eqref{eq:int2Vx} and taking into account Eq. \eqref{eq:Wpsi}, we obtain 
	  \begin{align}
	  \int\partial_{x}E dx  =\int\left\{\partial_{x}\psi\,\mathcal{G}\psi + \partial_{x}\eta\left[g\eta +\frac{1}{2}(\partial_{x}\psi)^2-\frac{\big(\mathcal{G}\psi+\partial_{x}\psi\partial_{x}\eta\big)^2}{2\left(1+
	  	|\partial_{x}\eta|^2\right)}\right]\right\}dx,
	  \end{align}
	  which is exactly Eq. \eqref{eq:IdE}, in view of the Hamiltonian Eqs. \eqref{ZCSHam}. 
	  \section{Vertical functions $Z_n$, $n\geq -2$}\label{App:Zn}
	  The vertical basis system $\{   Z_{n}   \} _{n\ge -2}$ is composed of two polynomial functions $\left\{Z_{-2} ,Z_{-1} \right\}$ and a set of eigenfunctions $\{   Z_{n}   \} _{n    \ge     0} $, normalized so that $Z_n(\eta)=1$:
	  \begin{subequations}\label{eq:Zadd}
	  		\begin{align}
	  		Z_{  -  2} (  z  ;  {\kern 1pt} \eta   ,  {\kern 1pt} h  )    &=    \frac{\mu _{  0}   h_{  0} +1}{2  {\kern 1pt} h_{  0} }   {\kern 1pt} \frac{(  z    +    h  )}{H} ^{{\kern 1pt} 2}       -      \frac{\mu _{  0}   h_{  0} +1}{2  h_{  0} }   {\kern 1pt} H      +      1,\label{eq:Zadda}\\
	  		Z_{  -  1} (  z  {\kern 1pt} ;  \eta   ,  h  )    &=    \frac{\mu _{  0}   h_{  0} -1}{2  h_{  0} }   {\kern 1pt} \frac{(  z    +    h  )}{H} ^{{\kern 1pt} 2}     +    {\kern 1pt} \frac{1}{h_{  0} }   {\kern 1pt} (  z    +    h  )    {\kern 1pt} -    {\kern 1pt} \frac{\mu _{  0}   h_{  0} +1}{2  h_{  0} }   {\kern 1pt} H    {\kern 1pt} +    {\kern 1pt} 1\label{eq:Zaddb}\\
	  		Z_{0}(  z  ;  {\kern 1pt} \eta   ,  {\kern 1pt} h  )     &=    \frac{\cosh (  k_{0} (z    +    h  ))}{\cosh (  k_{0}   H  )},\label{Zntrigc}\\
	  		Z_{n}(  z  ;  {\kern 1pt} \eta   ,  {\kern 1pt} h  )     &=    \frac{\cos (  k_{n} (z    +    h  ))}{\cos (  k_{n}   H  )},\quad   n    \ge     1, \label{Zntrigd}        
	  		\end{align}
	  	\end{subequations}
	  	In Eqs. \eqref{eq:Zadda}-\eqref{Zntrigd}, $H    =    H(  x  ,  t  )    =    \eta   (  x  ,  t  )    +    h  ( x  )$ is the local depth of the fluid, $\mu _{  0}   ,  {\kern 1pt} h_{  0} $ are two auxiliary constants and $k_{n}     =    k_{n}   (  x  ,  {\kern 1pt} t  )$, $n    \ge     0$ are the roots of the following transcendental equations:
	  \begin{subequations}\label{eq:kns}
	  	\begin{align}
	  	k_{  0} {\kern 1pt} H  {\kern 1pt} \tanh   (  k_{  0} {\kern 1pt} H  )    &=     \mu _{  0}   H(  x  ,  t  )\label{eq:kn0}\\
	  	k_{  n} {\kern 1pt} H  {\kern 1pt} \tan   (  k_{  n} {\kern 1pt} H  )    &=    -     \mu _{  0}   H(  x  ,  t  ),\,\,         \text{for}\,\,    n    \ge     1\label{eq:knn},
	  	\end{align}
	  \end{subequations}
that are efficiently solved at machine precision \citep{PPA18}. The auxiliary constant $h_{  0} $ is introduced only for dimensional purposes, and its value is taken to be a characteristic depth of the studied configuration, e.g. the depth in the incident wave region. The essential role of $\mu _{  0} $ is to formulate the free-surface boundary condition of the Sturm-Liouville problem defining the eigenfunctions $\{   Z_{n}   \} _{n    \ge     0} $ and it is chosen as $\mu_0=\omega_0^2/g$, where $\omega_0$ is the frequency of the incident wave. For more details we refer the reader to \citet{AB99,BA11,AP17semi}.

\section{Finite-Difference formulae}\label{App:FD}
For an arbitrary function $u(x)$ discretized on a regular grid of spacing $\delta x$, first and second derivatives in space are approximated by fourth-order finite difference formulae, as:
\begin{multline}\label{fd1}
(\partial_xu)_i =\\
\left\{
\begin{array}{ll}
\frac{1}{\delta x}\left(-\frac{25}{12}u_{i} + 4u_{i+1} - 3u_{i+2} + \frac{4}{3} u_{i+3} - \frac{1}{4} u_{i+4}\right),  &  i = 1 \\
\frac{1}{\delta x}\left(-\frac{1}{4}u_{i-1} - \frac{5}{6}u_{i} + \frac{3}{2}u_{i+1} - \frac{1}{2} u_{i+2} + \frac{1}{12} u_{i+3}\right),  &  i = 2 \\
\frac{1}{12\delta x}\left(u_{i-2}-8u_{i-1}+8u_{i+1}-u_{i+2}\right),&  i = 3,...,N_X-2 \\
\frac{1}{\delta x}\left(-\frac{1}{12}u_{i-3} + \frac{1}{2}u_{i-2} - \frac{3}{2}u_{i-1} + \frac{5}{6} u_{i} + \frac{1}{4} u_{i+1}\right),  &  i = N_X-1 \\
\frac{1}{\delta x}\left(\frac{1}{4}u_{i-4} - \frac{4}{3}u_{i-3} + 3u_{i-2} - 4 u_{i-1} + \frac{25}{12} u_{i}\right),  &  i = N_X
\end{array}
\right.
\end{multline}
\begin{multline}\label{fd2}
(\partial^2_xu)_i =\\
\left\{
\begin{array}{ll}
\frac{1}{\delta^2}\left(\frac{15}{4}u_{i} - \frac{77}{6}u_{i+1} + \frac{107}{6}u_{i+2} - 13 u_{i+3} + \frac{61}{12} u_{i+4} -\frac{5}{6} u_{i+5}\right),\quad\quad    i = 1 \\
\frac{1}{\delta^2}\left(\frac{5}{6}u_{i-1} - \frac{5}{4}u_{i} - \frac{1}{3}u_{i+1} + \frac{7}{6} u_{i+2} - \frac{1}{2} u_{i+3} + \frac{1}{12}u_{i+4}\right),\quad\quad\quad\quad    i = 2 \\
\frac{1}{12\delta^2}\left(-u_{i-2}+16 u_{i-1}-30u_i+16u_{i+1}-u_{i+2}\right),\quad\quad  i = 3,...,N_X-2 \\
\frac{1}{\delta^2}\left(\frac{5}{6}u_{i+1} - \frac{5}{4}u_{i} - \frac{1}{3}u_{i-1} + \frac{7}{6} u_{i-2} - \frac{1}{2} u_{i-3} + \frac{1}{12}u_{i-4}\right),\quad\,\,\,     i = N_X-1 \\
\frac{1}{\delta^2}\left(\frac{15}{4}u_{i} - \frac{77}{6}u_{i-1} + \frac{107}{6}u_{i-2} - 13 u_{i-3} + \frac{61}{12} u_{i-4} -\frac{5}{6} u_{i-5}\right),\quad\,\,\,    i = N_X
\end{array}
\right.
\end{multline}

		\singlespacing
		\bibliography{refs}

\begin{thebibliography}{67}
\expandafter\ifx\csname natexlab\endcsname\relax\def\natexlab#1{#1}\fi
\providecommand{\url}[1]{\texttt{#1}}
\providecommand{\href}[2]{#2}
\providecommand{\path}[1]{#1}
\providecommand{\DOIprefix}{doi:}
\providecommand{\ArXivprefix}{arXiv:}
\providecommand{\URLprefix}{URL: }
\providecommand{\Pubmedprefix}{pmid:}
\providecommand{\doi}[1]{\href{http://dx.doi.org/#1}{\path{#1}}}
\providecommand{\Pubmed}[1]{\href{pmid:#1}{\path{#1}}}
\providecommand{\bibinfo}[2]{#2}
\ifx\xfnm\relax \def\xfnm[#1]{\unskip,\space#1}\fi
\bibitem[{Athanassoulis and Belibassakis(1999)}]{AB99}
\bibinfo{author}{Athanassoulis, G.A.}, \bibinfo{author}{Belibassakis, K.A.},
  \bibinfo{year}{1999}.
\newblock \bibinfo{title}{A consistent coupled-mode theory for the propagation
  of small-amplitude water waves over variable bathymetry regions}.
\newblock \bibinfo{journal}{J. Fluid Mech.} \bibinfo{volume}{389},
  \bibinfo{pages}{275–301}.
\bibitem[{Athanassoulis et~al.(2017)Athanassoulis, Belibassakis and
  Papoutsellis}]{ABP17rogue}
\bibinfo{author}{Athanassoulis, G.A.}, \bibinfo{author}{Belibassakis, K.A.},
  \bibinfo{author}{Papoutsellis, {\relax Ch}.E.}, \bibinfo{year}{2017}.
\newblock \bibinfo{title}{{An exact Hamiltonian coupled-mode system with
  application to extreme design waves over variable bathymetry}}.
\newblock \bibinfo{journal}{J. Ocean Eng. Mar. Energy} \bibinfo{volume}{3},
  \bibinfo{pages}{373--383}.
\bibitem[{Athanassoulis and Papoutsellis(2015)}]{Ath15omae}
\bibinfo{author}{Athanassoulis, G.A.}, \bibinfo{author}{Papoutsellis, {\relax
  Ch}.E.}, \bibinfo{year}{2015}.
\newblock \bibinfo{title}{New form of the {H}amiltonian equations for the
  nonlinear water-wave problem, based on a new representation of the {D}t{N}
  operator, and some applications.}, in: \bibinfo{booktitle}{Proc. 34th Int.
  Conf. Ocean. Offshore Arct. Eng., ASME, St. John’s, Newfoundland, Canada}.
\newblock \bibinfo{note}{V007T06A029}.
\bibitem[{Athanassoulis and Papoutsellis(2017)}]{AP17semi}
\bibinfo{author}{Athanassoulis, G.A.}, \bibinfo{author}{Papoutsellis, {\relax
  Ch}.E.}, \bibinfo{year}{2017}.
\newblock \bibinfo{title}{Exact semi-separation of variables in waveguides with
  non-planar boundaries}.
\newblock \bibinfo{journal}{Proc. R. Soc. A} \bibinfo{volume}{473, 20170017}.
\bibitem[{Beji and Battjes(1993)}]{BB93}
\bibinfo{author}{Beji, S.}, \bibinfo{author}{Battjes, J.A.},
  \bibinfo{year}{1993}.
\newblock \bibinfo{title}{Experimental investigation of wave propagation over a
  bar}.
\newblock \bibinfo{journal}{Coast. Eng.} \bibinfo{volume}{19},
  \bibinfo{pages}{151--163}.
\bibitem[{Belibassakis and Athanassoulis(2011)}]{BA11}
\bibinfo{author}{Belibassakis, K.A.}, \bibinfo{author}{Athanassoulis, G.A.},
  \bibinfo{year}{2011}.
\newblock \bibinfo{title}{A coupled-mode system with application to nonlinear
  water waves propagating in finite water depth and in variable bathymetry
  regions}.
\newblock \bibinfo{journal}{Coast. Eng.} \bibinfo{volume}{58},
  \bibinfo{pages}{337--350}.
\bibitem[{Benjamin and Olver(1982)}]{BO82}
\bibinfo{author}{Benjamin, T.B.}, \bibinfo{author}{Olver, P.J.},
  \bibinfo{year}{1982}.
\newblock \bibinfo{title}{Hamiltonian structure, symmetries and conservation
  laws for water waves}.
\newblock \bibinfo{journal}{J. Fluid Mech.} \bibinfo{volume}{125},
  \bibinfo{pages}{137--185}.
\bibitem[{Bingham and Zhang(2007)}]{BZ07}
\bibinfo{author}{Bingham, H.B.}, \bibinfo{author}{Zhang, H.},
  \bibinfo{year}{2007}.
\newblock \bibinfo{title}{On the accuracy of finite difference solutions for
  nonlinear water waves}.
\newblock \bibinfo{journal}{J. Eng. Math.} \bibinfo{volume}{58},
  \bibinfo{pages}{211--228}.
\bibitem[{Bonneton et~al.(2011)Bonneton, Barthélemy, Chazel, Cienfuegos and
  Lannes}]{BBCCL11}
\bibinfo{author}{Bonneton, P.}, \bibinfo{author}{Barthélemy, E.},
  \bibinfo{author}{Chazel, F.}, \bibinfo{author}{Cienfuegos, R.},
  \bibinfo{author}{Lannes, D.}, \bibinfo{year}{2011}.
\newblock \bibinfo{title}{{Recent advances in Serre-Green Naghdi modelling for
  wave transformation, breaking and runup processes}}.
\newblock \bibinfo{journal}{Eur. J. Mech. B Fluids} \bibinfo{volume}{30},
  \bibinfo{pages}{589--597}.
\bibitem[{Briganti et~al.(2004)Briganti, Musumeci, Bellotti, Brocchini and
  Foti}]{B2004}
\bibinfo{author}{Briganti, R.}, \bibinfo{author}{Musumeci, R.E.},
  \bibinfo{author}{Bellotti, G.}, \bibinfo{author}{Brocchini, M.},
  \bibinfo{author}{Foti, E.}, \bibinfo{year}{2004}.
\newblock \bibinfo{title}{Boussinesq modeling of breaking waves: Description of
  turbulence}.
\newblock \bibinfo{journal}{J. Geophys. Res.} \bibinfo{volume}{109, C07015}.
\bibitem[{Brocchini(2013)}]{B13}
\bibinfo{author}{Brocchini, M.}, \bibinfo{year}{2013}.
\newblock \bibinfo{title}{{A reasoned overview on Boussinesq-type models: the
  interplay between physics, mathematics and numerics}}.
\newblock \bibinfo{journal}{Proc. R. Soc. A} \bibinfo{volume}{469(2160),
  20130496}.
\bibitem[{Broer(1974)}]{Broer74}
\bibinfo{author}{Broer, L.J.}, \bibinfo{year}{1974}.
\newblock \bibinfo{title}{{On the Hamiltonian theory of surface waves}}.
\newblock \bibinfo{journal}{Appl. Sci. Res} \bibinfo{volume}{30},
  \bibinfo{pages}{430--446}.
\bibitem[{Cao et~al.(1993)Cao, Beck and Schultz}]{CBS93}
\bibinfo{author}{Cao, Y.}, \bibinfo{author}{Beck, R.F.},
  \bibinfo{author}{Schultz, W.W.}, \bibinfo{year}{1993}.
\newblock \bibinfo{title}{An absorbing beach for numerical simulations of
  nonlinear waves in a wave tank.}, in: \bibinfo{booktitle}{8th Int. Workshop
  Water Waves and Floating Bodies}, pp. \bibinfo{pages}{17--20}.
\newblock \bibinfo{note}{St John's, Newfoundland, Canada}.
\bibitem[{do~Carmo et~al.(2019)do~Carmo, Ferreira and Pinto}]{CFP2019}
\bibinfo{author}{do~Carmo, J.S.A.}, \bibinfo{author}{Ferreira, J.},
  \bibinfo{author}{Pinto, L.}, \bibinfo{year}{2019}.
\newblock \bibinfo{title}{{On the accurate simulation of nearshore and dam
  break problems involving dispersive breaking waves}}.
\newblock \bibinfo{journal}{Wave Motion} \bibinfo{volume}{85},
  \bibinfo{pages}{125--143}.
\bibitem[{do~Carmo et~al.(2018)do~Carmo, Ferreira, Pinto and
  Romanazzi}]{CARMO2018}
\bibinfo{author}{do~Carmo, J.S.A.}, \bibinfo{author}{Ferreira, J.A.},
  \bibinfo{author}{Pinto, L.}, \bibinfo{author}{Romanazzi, G.},
  \bibinfo{year}{2018}.
\newblock \bibinfo{title}{{An improved Serre model: Efficient simulation and
  comparative evaluation}}.
\newblock \bibinfo{journal}{Applied Mathematical Modelling}
  \bibinfo{volume}{56}, \bibinfo{pages}{404 -- 423}.
\bibitem[{do~Carmo et~al.(1993)do~Carmo, Seabra-Santos and Almeida}]{Carmo93}
\bibinfo{author}{do~Carmo, J.S.A.}, \bibinfo{author}{Seabra-Santos, F.J.},
  \bibinfo{author}{Almeida, A.B.}, \bibinfo{year}{1993}.
\newblock \bibinfo{title}{{Numerical solution of the generalized Serre
  equations with the McCormack finite-difference scheme}}.
\newblock \bibinfo{journal}{Int. J. Numer. Methods Fluids}
  \bibinfo{volume}{16}, \bibinfo{pages}{725--738}.
\bibitem[{Chazel et~al.(2011)Chazel, Lannes and Marche}]{CLM11}
\bibinfo{author}{Chazel, F.}, \bibinfo{author}{Lannes, D.},
  \bibinfo{author}{Marche, F.}, \bibinfo{year}{2011}.
\newblock \bibinfo{title}{{Numerical simulation of strongly nonlinear and
  dispersive waves using a Green-Naghdi model}}.
\newblock \bibinfo{journal}{J. Sci. Comput.} \bibinfo{volume}{48},
  \bibinfo{pages}{105--116}.
\bibitem[{Cienfuegos et~al.(2010)Cienfuegos, Barthélemy and Bonneton}]{CRB10}
\bibinfo{author}{Cienfuegos, R.}, \bibinfo{author}{Barthélemy, E.},
  \bibinfo{author}{Bonneton, P.}, \bibinfo{year}{2010}.
\newblock \bibinfo{title}{{Wave-breaking model for Boussinesq-Type equations
  including roller effects in the mass conservation equation}}.
\newblock \bibinfo{journal}{J. Waterway, Port, Coastal, Ocean Eng.}
  \bibinfo{volume}{136}, \bibinfo{pages}{10--26}.
\bibitem[{Craig and Sulem(1993)}]{CS}
\bibinfo{author}{Craig, W.}, \bibinfo{author}{Sulem, C.}, \bibinfo{year}{1993}.
\newblock \bibinfo{title}{Numerical simulation of gravity waves}.
\newblock \bibinfo{journal}{J. Comp. Phys.} \bibinfo{volume}{108},
  \bibinfo{pages}{73--83}.
\bibitem[{Delestre et~al.(2013)Delestre, Lucas, Ksinant, Darboux, Laguerre,
  \relax{T.N. Tuoi Vo}, James and Cordier}]{NSWstoker_sol}
\bibinfo{author}{Delestre, O.}, \bibinfo{author}{Lucas, C.},
  \bibinfo{author}{Ksinant, P.A.}, \bibinfo{author}{Darboux, F.},
  \bibinfo{author}{Laguerre, C.}, \bibinfo{author}{\relax{T.N. Tuoi Vo}},
  \bibinfo{author}{James, F.}, \bibinfo{author}{Cordier, S.},
  \bibinfo{year}{2013}.
\newblock \bibinfo{title}{{SWASHES: a compilation of shallow water analytic
  solutions for hydraulic and environmental studies }}.
\newblock \bibinfo{journal}{Int. J. Numer. Meth. Fluids} \bibinfo{volume}{72},
  \bibinfo{pages}{269--300}.
\bibitem[{Derakhti et~al.(2016)Derakhti, Kirby, Shi and Ma}]{DKSM16}
\bibinfo{author}{Derakhti, M.}, \bibinfo{author}{Kirby, J.},
  \bibinfo{author}{Shi, F.}, \bibinfo{author}{Ma, G.}, \bibinfo{year}{2016}.
\newblock \bibinfo{title}{{Wave breaking in the surf zone and deep-water in a
  non-hydrostatic RANS model. Part 2: Turbulence and mean circulation}}.
\newblock \bibinfo{journal}{Ocean Modelling} \bibinfo{volume}{107},
  \bibinfo{pages}{139--150}.
\bibitem[{Dommermuth and Yue(1987)}]{DY87}
\bibinfo{author}{Dommermuth, D.G.}, \bibinfo{author}{Yue, D.K.P.},
  \bibinfo{year}{1987}.
\newblock \bibinfo{title}{A high-order spectral method for the study of
  nonlinear gravity waves.}
\newblock \bibinfo{journal}{J. Fluid Mech.} \bibinfo{volume}{184},
  \bibinfo{pages}{267 -- 288}.
\bibitem[{Gagarina et~al.(2014)Gagarina, Ambatia, van~der Vegt and
  Bokhove}]{GAVB14}
\bibinfo{author}{Gagarina, E.}, \bibinfo{author}{Ambatia, V.R.},
  \bibinfo{author}{van~der Vegt, J.J.W.}, \bibinfo{author}{Bokhove, O.},
  \bibinfo{year}{2014}.
\newblock \bibinfo{title}{{Variational space–time (dis)continuous Galerkin
  method for nonlinear free surface water waves }}.
\newblock \bibinfo{journal}{J. Comput. Phys.} \bibinfo{volume}{275},
  \bibinfo{pages}{459--483}.
\bibitem[{Gouin et~al.(2016)Gouin, Ducrozet and Ferrant}]{GDF2016}
\bibinfo{author}{Gouin, M.}, \bibinfo{author}{Ducrozet, G.},
  \bibinfo{author}{Ferrant, P.}, \bibinfo{year}{2016}.
\newblock \bibinfo{title}{Development and validation of a non-linear spectral
  model for water waves over variable depth}.
\newblock \bibinfo{journal}{Eur. J. Mech. B Fluids} \bibinfo{volume}{57},
  \bibinfo{pages}{115--128}.
\bibitem[{Grilli et~al.(2001)Grilli, Guyenne and Dias}]{GGD01}
\bibinfo{author}{Grilli, S.T.}, \bibinfo{author}{Guyenne, P.},
  \bibinfo{author}{Dias, F.}, \bibinfo{year}{2001}.
\newblock \bibinfo{title}{A fully non-linear model for three-dimensional
  overturning waves over an arbitrary bottom}.
\newblock \bibinfo{journal}{Int. J. Numer. Meth. Fluids} \bibinfo{volume}{35},
  \bibinfo{pages}{829--867}.
\bibitem[{Grilli et~al.(2019)Grilli, Horillo and Guignard}]{GG2019}
\bibinfo{author}{Grilli, S.T.}, \bibinfo{author}{Horillo, J.},
  \bibinfo{author}{Guignard, S.}, \bibinfo{year}{2019}.
\newblock \bibinfo{title}{Fully nonlinear potential flow simulations of wave
  shoaling over slopes: Spilling breaker model and integral wave properties}.
\newblock \bibinfo{journal}{Water Waves}
  \DOIprefix\doi{10.1007/s42286-019-00017-6}.
\bibitem[{Grilli and Horrillo(1997)}]{GH1997}
\bibinfo{author}{Grilli, S.T.}, \bibinfo{author}{Horrillo, J.},
  \bibinfo{year}{1997}.
\newblock \bibinfo{title}{{Numerical generation and absorption of fully
  nonlinear periodic waves}}.
\newblock \bibinfo{journal}{J. Eng. Mech.} \bibinfo{volume}{123},
  \bibinfo{pages}{1060--1069}.
\bibitem[{Guignard and Grilli(2001)}]{GG2001}
\bibinfo{author}{Guignard, S.}, \bibinfo{author}{Grilli, S.T.},
  \bibinfo{year}{2001}.
\newblock \bibinfo{title}{{Modeling of wave shoaling in a 2D-NWT using a
  spilling breaker model}}, in: \bibinfo{booktitle}{Proc. ISOPE 2001 Conf.}
\newblock \bibinfo{note}{Stavanger, Norway, ISOPE-I-01-253}.
\bibitem[{Guyenne and Nicholls(2007)}]{GN07}
\bibinfo{author}{Guyenne, P.}, \bibinfo{author}{Nicholls, D.},
  \bibinfo{year}{2007}.
\newblock \bibinfo{title}{A high-order spectral method for nonlinear water
  waves over moving bottom topography}.
\newblock \bibinfo{journal}{SIAM J. Sci. Comput} \bibinfo{volume}{30},
  \bibinfo{pages}{81--101}.
\bibitem[{Heitner and Housner(1970)}]{HH70}
\bibinfo{author}{Heitner, K.}, \bibinfo{author}{Housner, G.},
  \bibinfo{year}{1970}.
\newblock \bibinfo{title}{Numerical model for tsunami run-up}.
\newblock \bibinfo{journal}{Coast. Eng. Div.} \bibinfo{volume}{96},
  \bibinfo{pages}{701--719}.
\bibitem[{Karambas and Koutitas(1992)}]{KK92}
\bibinfo{author}{Karambas, T.V.}, \bibinfo{author}{Koutitas, C.},
  \bibinfo{year}{1992}.
\newblock \bibinfo{title}{{A breaking wave propagation model based on the
  Boussinesq equations}}.
\newblock \bibinfo{journal}{Coast. Eng.} \bibinfo{volume}{18},
  \bibinfo{pages}{1--19}.
\bibitem[{Karambas and Memos(2009)}]{CM2009}
\bibinfo{author}{Karambas, T.V.}, \bibinfo{author}{Memos, C.D.},
  \bibinfo{year}{2009}.
\newblock \bibinfo{title}{{Boussinesq model for weakly nonlinear fully
  dispersive water waves}}.
\newblock \bibinfo{journal}{J. Waterway, Port, Coastal, Ocean Eng.}
  \bibinfo{volume}{135}, \bibinfo{pages}{187--199}.
\bibitem[{Kazolea et~al.(2014)Kazolea, Delis and Synolakis}]{KDS14}
\bibinfo{author}{Kazolea, M.}, \bibinfo{author}{Delis, A.I.},
  \bibinfo{author}{Synolakis, C.E.}, \bibinfo{year}{2014}.
\newblock \bibinfo{title}{{Numerical treatment of wave breaking on unstructured
  finite volume approximations for extended Boussinesq-type equations}}.
\newblock \bibinfo{journal}{J. Comp. Phys.} \bibinfo{volume}{271},
  \bibinfo{pages}{281--305}.
\bibitem[{Kazolea and Ricchiuto(2018)}]{KR18}
\bibinfo{author}{Kazolea, M.}, \bibinfo{author}{Ricchiuto, M.},
  \bibinfo{year}{2018}.
\newblock \bibinfo{title}{{On wave breaking for Boussinesq-type models}}.
\newblock \bibinfo{journal}{Ocean Modelling} \bibinfo{volume}{123},
  \bibinfo{pages}{16--39}.
\bibitem[{Kennedy et~al.(2000)Kennedy, Chen, Kirby and
  Dalrymple}]{Kennedyetal2000}
\bibinfo{author}{Kennedy, A.B.}, \bibinfo{author}{Chen, Q.C.},
  \bibinfo{author}{Kirby, J.T.}, \bibinfo{author}{Dalrymple, R.},
  \bibinfo{year}{2000}.
\newblock \bibinfo{title}{{Boussinesq modeling of wave transformation,
  breaking, and runup I:1D }}.
\newblock \bibinfo{journal}{J. Waterway, Port, Coastal, Ocean Eng.}
  \bibinfo{volume}{126}, \bibinfo{pages}{39--47}.
\bibitem[{Kirby(2016)}]{K16}
\bibinfo{author}{Kirby, J.T.}, \bibinfo{year}{2016}.
\newblock \bibinfo{title}{{Boussinesq models and their application to coastal
  processes across a wide range of scales}}.
\newblock \bibinfo{journal}{J. Waterway, Port, Coastal, Ocean Eng.}
  \bibinfo{volume}{142(6), 03116005}.
\bibitem[{Kurnia and van Groesen(2014)}]{KvG14}
\bibinfo{author}{Kurnia, R.}, \bibinfo{author}{van Groesen, E.},
  \bibinfo{year}{2014}.
\newblock \bibinfo{title}{High order hamiltonian water wave models with
  wave-breaking mechanism}.
\newblock \bibinfo{journal}{Coast. Eng.} \bibinfo{volume}{93},
  \bibinfo{pages}{55--70}.
\bibitem[{Lannes(2013)}]{Lannesb}
\bibinfo{author}{Lannes, D.}, \bibinfo{year}{2013}.
\newblock \bibinfo{title}{{The Water Waves Problem. Mathematical Analysis and
  Asymptotics}}. volume \bibinfo{volume}{188} of
  \textit{\bibinfo{series}{Mathematical Surveys and Monographs}}.
\newblock \bibinfo{publisher}{American Mathematical Society}.
\bibitem[{Lannes and Bonneton(2009)}]{LB09}
\bibinfo{author}{Lannes, D.}, \bibinfo{author}{Bonneton, P.},
  \bibinfo{year}{2009}.
\newblock \bibinfo{title}{Derivation of asymptotic two-dimensional
  time-dependent equations for surface water wave propagation}.
\newblock \bibinfo{journal}{Phys. Fluids} \bibinfo{volume}{21, 016601}.
\bibitem[{Madsen et~al.(2006)Madsen, Fuhrman and Wang}]{MFW2006}
\bibinfo{author}{Madsen, P.A.}, \bibinfo{author}{Fuhrman, D.R.},
  \bibinfo{author}{Wang, B.}, \bibinfo{year}{2006}.
\newblock \bibinfo{title}{{A Boussinesq-type method for fully nonlinear waves
  interacting with a rapidly varying bathymetry}}.
\newblock \bibinfo{journal}{Coast. Eng.} \bibinfo{volume}{53},
  \bibinfo{pages}{487--504}.
\bibitem[{Madsen et~al.(1997)Madsen, Sorensen and Schäffer}]{MSS97}
\bibinfo{author}{Madsen, P.A.}, \bibinfo{author}{Sorensen, O.R.},
  \bibinfo{author}{Schäffer, H.A.}, \bibinfo{year}{1997}.
\newblock \bibinfo{title}{Surf zone dynamics simulated by a boussinesq type
  model. part i. model description and cross-shore motion of regular waves}.
\newblock \bibinfo{journal}{Coast. Eng.} \bibinfo{volume}{32},
  \bibinfo{pages}{255--287}.
\bibitem[{Milder(1990)}]{Milder90}
\bibinfo{author}{Milder, D.M.}, \bibinfo{year}{1990}.
\newblock \bibinfo{title}{{The effects of truncation on surface-wave
  Hamiltonians}}.
\newblock \bibinfo{journal}{J. Fluid Mech.} \bibinfo{volume}{216},
  \bibinfo{pages}{249--262}.
\bibitem[{Miles(1977)}]{Miles}
\bibinfo{author}{Miles, J.W.}, \bibinfo{year}{1977}.
\newblock \bibinfo{title}{{On Hamilton's principle for surface waves}}.
\newblock \bibinfo{journal}{J. Fluid Mech} \bibinfo{volume}{83},
  \bibinfo{pages}{153--158}.
\bibitem[{Mitsotakis et~al.(2017)Mitsotakis, Dutykh and Carter}]{MITSOT2017}
\bibinfo{author}{Mitsotakis, D.}, \bibinfo{author}{Dutykh, D.},
  \bibinfo{author}{Carter, J.}, \bibinfo{year}{2017}.
\newblock \bibinfo{title}{{On the nonlinear dynamics of the traveling-wave
  solutions of the Serre system}}.
\newblock \bibinfo{journal}{Wave Motion} \bibinfo{volume}{70},
  \bibinfo{pages}{166 -- 182}.
\bibitem[{Mitsotakis et~al.(2015)Mitsotakis, Synolakis and McGuiness}]{MSM15}
\bibinfo{author}{Mitsotakis, D.}, \bibinfo{author}{Synolakis, C.E.},
  \bibinfo{author}{McGuiness, M.}, \bibinfo{year}{2015}.
\newblock \bibinfo{title}{{A modified Galerkin/Finite Element Method for the
  numerical solution of the Serre-Green-Nagdhi system}}.
\newblock \bibinfo{journal}{Int. J. Numer. Meth.} \bibinfo{volume}{83(10)},
  \bibinfo{pages}{755--778}.
\bibitem[{Nwogu(1993)}]{Nwogu93}
\bibinfo{author}{Nwogu, O.G.}, \bibinfo{year}{1993}.
\newblock \bibinfo{title}{{Alternative form of Boussinesq equations for
  nearshore wave propagation}}.
\newblock \bibinfo{journal}{J. Waterway, Port, Coastal, Ocean Eng.}
  \bibinfo{volume}{119}, \bibinfo{pages}{618--638}.
\bibitem[{Papathanasiou et~al.(2018)Papathanasiou, Papoutsellis and
  Athanassoulis}]{PPA18}
\bibinfo{author}{Papathanasiou, T.}, \bibinfo{author}{Papoutsellis, {\relax
  Ch}.E.}, \bibinfo{author}{Athanassoulis, G.A.}, \bibinfo{year}{2018}.
\newblock \bibinfo{title}{{Semi-explicit solutions to the water-wave dispersion
  relation and their role in the non-linear Hamiltonian coupled-mode theory}}.
\newblock \bibinfo{journal}{J. Eng. Math.} \bibinfo{volume}{114(1)},
  \bibinfo{pages}{87–114}.
\bibitem[{Papoutsellis(2016)}]{Papoutsellis}
\bibinfo{author}{Papoutsellis, {\relax Ch}.E.}, \bibinfo{year}{2016}.
\newblock \bibinfo{title}{Nonlinear water waves over varying bathymetry:
  theoretical and numerical study using variational methods}.
\newblock Ph.D. thesis. National Technical University of Athens.
\newblock
  \bibinfo{note}{\url{http://dspace.lib.ntua.gr/handle/123456789/44741}}.
\bibitem[{Papoutsellis and Athanassoulis(2017)}]{PA17arxiv}
\bibinfo{author}{Papoutsellis, {\relax Ch}.E.}, \bibinfo{author}{Athanassoulis,
  G.A.}, \bibinfo{year}{2017}.
\newblock \bibinfo{title}{{A new efficient Hamiltonian approach to the
  nonlinear water-wave problem over arbitrary bathymetry}}
  \bibinfo{note}{\url{http://arxiv.org/abs/1704.03276}}.
\bibitem[{Papoutsellis et~al.(2018)Papoutsellis, Charalampopoulos and
  Athanassoulis}]{PCA18}
\bibinfo{author}{Papoutsellis, {\relax Ch}.E.},
  \bibinfo{author}{Charalampopoulos, A.}, \bibinfo{author}{Athanassoulis,
  G.A.}, \bibinfo{year}{2018}.
\newblock \bibinfo{title}{{Implementation of a fully nonlinear Hamiltonian
  Coupled-Mode Theory, and application to solitary wave problems over
  bathymetry}}.
\newblock \bibinfo{journal}{Eur. J. Mech. B Fluids} \bibinfo{volume}{72},
  \bibinfo{pages}{199--224}.
\bibitem[{Raoult et~al.(2016)Raoult, Benoit and Yates}]{RBY16}
\bibinfo{author}{Raoult, C.}, \bibinfo{author}{Benoit, M.},
  \bibinfo{author}{Yates, M.L.}, \bibinfo{year}{2016}.
\newblock \bibinfo{title}{Validation of a fully nonlinear and dispersive wave
  model with laboratory non-breaking experiments}.
\newblock \bibinfo{journal}{Coast. Eng.} \bibinfo{volume}{114},
  \bibinfo{pages}{194--207}.
\bibitem[{Raoult et~al.(2019)Raoult, Benoit and Yates}]{RBY19}
\bibinfo{author}{Raoult, C.}, \bibinfo{author}{Benoit, M.},
  \bibinfo{author}{Yates, M.L.}, \bibinfo{year}{2019}.
\newblock \bibinfo{title}{{Development and validation of a 3D RBF-spectral
  model for coastal wave simulation}}.
\newblock \bibinfo{journal}{J. Comput. Phys.} \bibinfo{volume}{378},
  \bibinfo{pages}{278--302}.
\bibitem[{Roeber et~al.(2010)Roeber, Cheung and Kobayashi}]{RCK10}
\bibinfo{author}{Roeber, V.}, \bibinfo{author}{Cheung, K.},
  \bibinfo{author}{Kobayashi, M.}, \bibinfo{year}{2010}.
\newblock \bibinfo{title}{{Shock-capturing Boussinesq-type model for nearshore
  wave processes}}.
\newblock \bibinfo{journal}{Coast. Eng.} \bibinfo{volume}{57},
  \bibinfo{pages}{407--423}.
\bibitem[{Sch\"{a}ffer et~al.(1993)Sch\"{a}ffer, Madsen and Deigaard}]{SMD93}
\bibinfo{author}{Sch\"{a}ffer, H.}, \bibinfo{author}{Madsen, P.},
  \bibinfo{author}{Deigaard, R.}, \bibinfo{year}{1993}.
\newblock \bibinfo{title}{{A Boussinesq model for waves breaking in shallow
  water}}.
\newblock \bibinfo{journal}{Coast. Eng.} \bibinfo{volume}{20},
  \bibinfo{pages}{185--202}.
\bibitem[{Seiffert and Ducrozet(2017)}]{SD17}
\bibinfo{author}{Seiffert, B.R.}, \bibinfo{author}{Ducrozet, G.},
  \bibinfo{year}{2017}.
\newblock \bibinfo{title}{{Simulation of breaking waves using the high-order
  spectral method with laboratory experiments: wave-breaking energy
  dissipation}}.
\newblock \bibinfo{journal}{Ocean Dynam.} \bibinfo{volume}{68},
  \bibinfo{pages}{65--89}.
\bibitem[{Stoker(1957)}]{St}
\bibinfo{author}{Stoker, J.J.}, \bibinfo{year}{1957}.
\newblock \bibinfo{title}{{Water Waves: The Mathematical Theory with
  Applications}}. volume~\bibinfo{volume}{4} of \textit{\bibinfo{series}{Pure
  and Applied Mathematics}}.
\newblock \bibinfo{publisher}{Interscience Publishers}, \bibinfo{address}{New
  York}.
\bibitem[{Svendsen(1984)}]{Sv84}
\bibinfo{author}{Svendsen, I.A.}, \bibinfo{year}{1984}.
\newblock \bibinfo{title}{Mass flux and undertow in a surf zone}.
\newblock \bibinfo{journal}{Coast. Eng.} \bibinfo{volume}{8},
  \bibinfo{pages}{374--365}.
\bibitem[{Tian and Sato(2008)}]{TS08}
\bibinfo{author}{Tian, Y.}, \bibinfo{author}{Sato, S.}, \bibinfo{year}{2008}.
\newblock \bibinfo{title}{A numerical model on the interaction between
  nearshore nonlinear waves and strong currents}.
\newblock \bibinfo{journal}{Coast. Eng. J.} \bibinfo{volume}{50},
  \bibinfo{pages}{369--395}.
\bibitem[{Tian et~al.(2010)Tian, Perlin and Choi}]{TPC10}
\bibinfo{author}{Tian, Z.}, \bibinfo{author}{Perlin, M.},
  \bibinfo{author}{Choi, W.}, \bibinfo{year}{2010}.
\newblock \bibinfo{title}{{Energy dissipation in two-dimensional unsteady
  plunging breakers and an eddy viscosity model}}.
\newblock \bibinfo{journal}{J. Fluid Mech.} \bibinfo{volume}{655},
  \bibinfo{pages}{217--257}.
\bibitem[{Ting and Kirby(1994)}]{TK94}
\bibinfo{author}{Ting, F.C.K.}, \bibinfo{author}{Kirby, J.T.},
  \bibinfo{year}{1994}.
\newblock \bibinfo{title}{Observation of undertow and turbulence in a
  laboratory surf zone}.
\newblock \bibinfo{journal}{Coast. Eng.} \bibinfo{volume}{24},
  \bibinfo{pages}{51--80}.
\bibitem[{Tissier et~al.(2012)Tissier, Bonneton, Marche, Chazel and
  Lannes}]{TBMCL12}
\bibinfo{author}{Tissier, M.}, \bibinfo{author}{Bonneton, P.},
  \bibinfo{author}{Marche, F.}, \bibinfo{author}{Chazel, F.},
  \bibinfo{author}{Lannes, D.}, \bibinfo{year}{2012}.
\newblock \bibinfo{title}{{A new approach to handle wave breaking in fully
  non-linear Boussinesq models}}.
\newblock \bibinfo{journal}{Coast. Eng.} \bibinfo{volume}{67},
  \bibinfo{pages}{54--66}.
\bibitem[{Tonelli and Petti(2009)}]{TP09}
\bibinfo{author}{Tonelli, M.}, \bibinfo{author}{Petti, M.},
  \bibinfo{year}{2009}.
\newblock \bibinfo{title}{{Hybrid finite-volume finite-difference scheme for
  2DH improved Boussinesq equations}}.
\newblock \bibinfo{journal}{Coast. Eng.} \bibinfo{volume}{56},
  \bibinfo{pages}{609--620}.
\bibitem[{Veeramony and Svendsen(2000)}]{VS00}
\bibinfo{author}{Veeramony, J.}, \bibinfo{author}{Svendsen, I.A.},
  \bibinfo{year}{2000}.
\newblock \bibinfo{title}{The flow in surf-zone waves}.
\newblock \bibinfo{journal}{Coast. Eng.} \bibinfo{volume}{39},
  \bibinfo{pages}{93--122}.
\bibitem[{Yates and Benoit(2015)}]{YB15}
\bibinfo{author}{Yates, M.L.}, \bibinfo{author}{Benoit, M.},
  \bibinfo{year}{2015}.
\newblock \bibinfo{title}{Accuracy and efficiency of two numerical methods of
  solving the potential flow problem for highly nonlinear and dispersive water
  waves}.
\newblock \bibinfo{journal}{Int. J. Numer. Methods Fluids}
  \bibinfo{volume}{77}, \bibinfo{pages}{616--640}.
\bibitem[{Zakharov(1968)}]{Zak}
\bibinfo{author}{Zakharov, V.E.}, \bibinfo{year}{1968}.
\newblock \bibinfo{title}{Stability of periodic waves of finite amplitude on
  the surface of a deep fluid}.
\newblock \bibinfo{journal}{J. Appl. Mech. Tech. Phys} \bibinfo{volume}{9},
  \bibinfo{pages}{86--94}.
\bibitem[{Zelt(1991)}]{Z91}
\bibinfo{author}{Zelt, J.A.}, \bibinfo{year}{1991}.
\newblock \bibinfo{title}{The run-up of nonbreaking and breaking solitary
  waves}.
\newblock \bibinfo{journal}{Coast. Eng.} \bibinfo{volume}{15},
  \bibinfo{pages}{205--246}.
\bibitem[{Zhao et~al.(2014)Zhao, Duan and Ertekin}]{ZDE14}
\bibinfo{author}{Zhao, B.B.}, \bibinfo{author}{Duan, W.Y.},
  \bibinfo{author}{Ertekin, R.C.}, \bibinfo{year}{2014}.
\newblock \bibinfo{title}{{Application of higher-level GN theory to some wave
  transformation problems}}.
\newblock \bibinfo{journal}{Coast. Eng.} \bibinfo{volume}{83},
  \bibinfo{pages}{177--189}.

\end{thebibliography}

\end{document}